\documentclass{osa-article}

\journal{osac}
\usepackage{xcolor}
\articletype{Research Article}
\usepackage{lineno}
\usepackage{subcaption}
\newcommand{\rd}{\mathrm{d}}

\newcommand{\calH}{\mathcal{H}}
\newcommand{\calR}{\mathcal{R}}
\newcommand{\rmf}{\mathrm{f}}
\newcommand{\EE}{\mathbb{E}}

\newcommand{\revision}[1]{{\color{black}{#1}}}
\begin{document}

%\title{Optimal beam calculations in weak turbulence}
\title{Computation of optimal beams in weak turbulence}

\author{Qin Li,\authormark{1} Anjali Nair,\authormark{1,*} and Samuel N Stechmann\authormark{1,2}}

\address{\authormark{1}Department of Mathematics, University of Wisconsin - Madison, Madison, WI 53706 USA\\
\authormark{2}Department of Atmospheric and Oceanic Sciences, University of Wisconsin - Madison, Madison, WI 53706 USA}

\email{\authormark{*}nair25@wisc.edu} %% email address is required



\vspace{12pt}
\noindent
Submitted on August 9, 2022

\begin{abstract}
When an optical beam propagates through a turbulent medium such as the atmosphere or ocean, the beam will become distorted. It is then natural to seek the best or optimal beam that is distorted least, under some metric such as intensity or scintillation. We seek to maximize the light intensity at the receiver using the paraxial wave equation with weak-fluctuation as the model. In contrast to  classical results that typically confine original laser beams to be from a special class, we allow the beam to be general, which leads to an eigenvalue problem of a large-sized matrix with each entry being a multi-dimensional integral. This is an expensive and sometimes infeasible computational task in many practically reasonable settings. To overcome this expense, \revision{in a change from past calculations of optimal beams, we transform the calculation from physical space to Fourier space. Since the structure of the turbulence is commonly described in Fourier space, the computational cost is significantly reduced. This also allows us to incorporate some optional turbulence assumptions, such as homogeneous-statistics assumption, small-length-scale cutoff assumption, and Markov assumption, to further reduce the dimension of the numerical integral.} The proposed methods provide a computational strategy that is numerically feasible, and results are demonstrated in several numerical examples. \revision{These results provide further evidence that special beams can be defined to have beam divergence that is small.}
%and accurate for practical problems.
% \\ \\
% \textcolor{red}{Keywords? waves, random medium, turbulent fluctuations,
% atmospheric turbulence, ocean turbulence, laser communications,
% free-space optical communications, ... ?}
\end{abstract}

\section{Introduction}

When optical beams propagate through a random medium, they are subject to distortions that lead to unwanted phenomena like intensity reduction and scintillation~\cite{strohbehn1978laser, andrews2005laser}. It is then often desirable to search for beams that are optimal under certain criteria.

% \ql{Please check and fill in: The two most prominent criteria are to maximize light intensity, and to minimize the scintillation index, both at the receiver. There is a vast literature on the subject taking either the computational or experimental perspective. For example, in~\cite{}, the authors have conducted a series of experimental works to evaluate the scintillation index with various possible beam in different environmental setups. In~\cite{}, the authors have computationally deployed XX methods.}

The criterion taken in this paper is to maximize light intensity at the receiver. Mathematically it has been proved to be a very clean problem. It was shown that the optimal beam under this criterion has to be coherent~\cite{schulz2004iterative,schulz2005optimal}. Moreover, it was proved that this optimal, coherent beam is associated with the largest eigenvalue of a matrix or operator:
\begin{equation}\label{eqn:calH}
   \calH=\EE[H]\,,\quad\text{with}\quad H(X_1,X_2)=\int\limits_{X'\in\mathcal{R}}h(X_1,X')h^\ast(X_2,X')\mathrm{d}X'\,.
\end{equation}
Here $h(X_1,X')$ is the propagator (Green's function) that propagates light from the origin $X_1$ to $X'$. $\calR$ is the receiver region, and the expected value $\EE$ takes average over all possible configurations of the random medium. By definition, $H(X_1,X_2)$ is a Hermitian kernel with non-negative eigenvalues. Since the task of finding an optimal beam is equivalent to the task of finding the eigenvector or eigenfunction associated with the largest eigenvalue of $\mathcal{H}=\EE[H]$, what remains is to mathematically and computationally formulate $\mathcal{H}$.

This turns out to be a very challenging task in practice. There are two obstacles. First, the expected value $\EE$ means all random media configurations need to be taken into account. Numerically, suppose we utilize Monte Carlo to sample many representative configurations; then a different $H$ needs to be evaluated for each of these samples. Second, computing each $H$ is already a major challenge. It amounts to evaluating all Green's functions $h$ that maps every $X\in\mathcal{A}$, the aperture, to every $X'\in\mathcal{R}$. This is a drastic cost for a problem in three-dimensional space, $\mathbb{R}^3$.

% for every random media configuration, one needs to evaluate, for this particular media, all possible $h$s, the Green's function that maps every $X\in\mathcal{A}$ to every $X'\in\mathcal{R}$. Secondly, one needs to take an expectation $\EE$ over all random media configurations. While calculating $h$ requires numerical fine discretization that leads to high degrees of freedom and prohibitive computational cost, taking the ensemble $\EE$ furthermore requires such process carried out repeatedly for many rounds of configurations (as in Monte Carlo sampling). These together make the problem numerically infeasible.

These difficulties have made it practically impossible to find the optimal beams. To overcome the difficulties, \revision{we consider the following context and approach. As a first aspect, we consider} perturbation theory in the weak fluctuation regime. One convenience of studying the weak fluctuation regime is that the nonlinear dependence on the randomness is now linearized around the deterministic setup. Instead of sampling many random configurations of the media, and computing all the associated Green's functions before taking the expected value $\EE$ over the media randomness, now one can find an analytical expression, and the expected value $\EE$ gets directly applied to the medium. Eventually it is the moments of the medium's randomness that enter into the analytical expression, and, if desired, empirical models such as the Kolmogorov spectrum can be employed in a straightforward manner. While analytical expressions for the moments of the solution to the \revision{Paraxial Wave Equation (PWE)} have been computed before using perturbation theory, they have been for specific classes of sources like Gaussian beams~\cite{andrews2005laser}. Here we compute the second moment for general source functions for general weakly turbulent models, and use this to compute optimal beams. 

However, such a formulation of the analytical expression for $\mathcal{H}$ is still hard to compute. The fine discretization and the high dimensionality issue encountered in the nonlinear regime are still present here. As will be presented in Section~\ref{sec:calH_physical}, this expression results in an $8$-fold integral in a full 3D simulation for each entry of $\mathcal{H}$, and is numerically prohibitive. \revision{As some intuition toward reducing the computational cost, one observation is that many empirical models for atmospheric turbulence have special structure on the Fourier domain, and thus it is expected that calculations can be significantly simplified if conducted on the Fourier side. The Fourier-space perspective, however, has not been taken previously in the literature on optimal beams. In our calculation, we fully take advantage of the encoded turbulence structure in the Fourier domain, and convert our calculation to that of $\calH_\rmf$, the Fourier-space counterpart of $\mathcal{H}$; see Section~\ref{sec:calH_Fourier}.} This allows us to incorporate some well-known assumptions on the medium's structure with ease. We use homogeneous assumption, small-length-scale cutoff assumption, and Markov assumption sequentially, and observe the reduction from an $8$-fold integral, to a $6$-fold, to a $2$-fold and finally to a $1$-fold integral respectively, making it gradually more and more numerically feasible, see Section~\ref{sec:assumptions}. Some properties of $\calH$ and $\calH_\rmf$ are shown in Section~\ref{sec:properties}, and numerical evidences are provided in Section~\ref{sec:numerics}.

The methods here provide an additional contribution to the literature on designing optimal beams in the presence of turbulence.  
As mentioned above,
a mathematical formulation for optimal beams was given in
~\cite{schulz2004iterative,schulz2005optimal}. 
Few numerical examples of optimal beams have been presented.
In the present paper, we find optimal beams for the
paraxial wave equation in the weak turbulence setting
where the refractive index is allowed to be a general random function
as a representation of turbulent fluctuations.
Past work
has investigated optimal beams under other frameworks 
that introduce additional assumptions, such as the
phase screen model and the extended Huygens--Fresnel (eHF)
principle \cite{liu2006optimal,zhou2018optimal,slepian1964prolate, belmonte2018approaching, shapiro2005ultimate},
or searching within a special class of beams such as
the Gaussian Schell-model beams
\cite{voelz2009metric}.

More broadly, the problem of finding optimal beams is
motivated by much past work on increasing beam intensity
or reducing scintillation. From past work on these topics,
it is generally known that coherent beams maximize the received intensity, although coherent beams do not perform well with respect to other metrics like scintillation; indeed, turbulence has a degrading effect on many types of beams~\cite{schulz2005optimal,cai2006average,polynkin2007optimized,qian2009numerical,korotkova2014scintillation,borcea2020multimode}. 
To reduce the effects of scintillation, past work has
identified the importance of partially coherent beams
\cite{dogariu2003propagation,chen2008propagation, ji2008spreading, qian2009numerical, polynkin2007optimized,gbur2002spreading,korotkova2004model,ricklin2003atmospheric}.
The problem of minimizing scintillation is more challenging
from a computational perspective, and we therefore focus here
on the more tractable problem of maximizing intensity.

\section{Problem setup and notation}
Consider a laser beam source \revision{with optical field }$\phi$ that is located in a transmitter region $\mathcal{A}$, and let $U$ denote the field received at the receiver region $\mathcal{R}$. For waves that are governed by a linear equation, $U$ can be represented as:
\begin{equation}\label{eqn:U}
U(X')=\int\limits_{X\in\mathcal{A}}h_\omega(X,X')\phi(X)\mathrm{d}X\,.
\end{equation}
This $h_\omega(X,X')$ is termed the propagator that sends light information from the aperture location $X$ to the receiver location $X'$, so it naturally includes the medium information. This propagator depends on the specific configuration of the media, as indicated by the subindex $\omega$. We surpress this subindex for the conciseness of the notation when the context is clear.

There are two sources of randomness. The source $\phi$ is generated by the laser and contains random fluctuation, and the medium also presents turbulence~\cite{gbur2014partially}. When a specific source and medium configuration is fixed, and the light intensity can be calculated as
\begin{align}
    I_i(X')=|U(X')|^2=\int_{X_{1},X_2\in\mathcal{A}}h(X_1,X')\phi(X_1)h^\ast(X_2,X')\phi^\ast(X_2)\mathrm{d}X_1\mathrm{d}X_2,\quad X'\in\mathcal{R}\, ,
\end{align}
where we use superscript $^*$ to denote the complex conjugate. Taking the average with respect to both the source fluctuations and media turbulence, we have the averaged intensity:
\begin{equation}
    I(X')=\mathbb{E}\Big[\int\limits_{X_{1}\in\mathcal{A}}\int\limits_{X_2\in\mathcal{A}}h(X_1,X')J(X_1,X_2)h^\ast(X_2,X')\mathrm{d}X_1\mathrm{d}X_2\Big],\quad X'\in\mathcal{R}\,,
\end{equation}
where $J(X_1,X_2)=\langle\phi(X_1)\phi^\ast(X_2)\rangle$ is the mutual intensity function of the source. Throughout the paper we use $\langle\cdot\rangle$ to denote the averaging with respect to the source randomness, and $\mathbb{E}[\cdot]$ to denote the averaging with respect to the randomness in the medium.

Designing optimal laser beam amounts to tuning $J$ that achieves the highest intensity. To do so, the intensity maximization problem is formulated as%We are interested in maximizing the total average intensity at the receiver, given a fixed total transmitted intensity $I_0$ (say, equal to 1) i.e, 
\begin{equation}\label{eqn:optimization}
    \max_{J} I_\mathcal{R}=\max_{J}\int\limits_{X'\in\mathcal{R}}I(X')\mathrm{d}X'\,,\quad\text{subject to}\quad I_0 = 1\,,
\end{equation}
where $I_0=\int_{X\in\mathcal{A}}J(X,X)\mathrm{d}X=\int_{X\in\mathcal{A}}\langle\phi(X)\phi^\ast(X)\rangle\mathrm{d}X$ 
is the light intensity at the initial state.

A useful reformulation of \eqref{eqn:optimization} follows from noting that, by definition, $J$ is an Hermitian matrix. Hence, following earlier work \cite{wolf1982new}, we can write $J$ in terms of its coherent mode expansion as
\begin{equation}
    J(X_1,X_2)=\sum\limits_k\alpha_k\psi_k(X_1)\psi_k^\ast(X_2), \quad X_1,X_2\in\mathcal{A},
\end{equation}
where $\{\alpha_k\}$ are non-negative weights and $\{\psi_k\}$ is a set of functions that are orthonormal over the transmitter region. Considering all candidates for $J$ is then equivalent to considering all $\{\alpha_k,\psi_k\}$ pairs~\cite{gori1983mode,shirai2003mode,habashy1997application}. This formulation allows us to rewrite the constraint as
\begin{equation}\label{eqn:I0}
    I_0=\sum_k\alpha_k=1\,,
\end{equation}
and the objective function $I_\calR$ becomes:
% The coherent mode representation is widely used to provide simplified computations for beam propagation ~\cite{gori1983mode,shirai2003mode}. This approach has also been used to study inverse problems ~\cite{habashy1997application}. In terms of the coherent mode expansion, the transmitted intensity is given by
% and the total average received intensity by
\begin{equation}\label{eqn:I}
    I_\mathcal{R}=\sum\limits_k\alpha_k\int\limits_{X_1\in\mathcal{A}}\int\limits_{X_2\in\mathcal{A}}\psi_k(X_1)\EE[H(X_1,X_2)]\psi_k^\ast(X_2)\mathrm{d}X_1\mathrm{d}X_2\,,
\end{equation}
where for all $X_1,X_2\in\mathcal{A}$ we use the same notation as in~\eqref{eqn:calH} and set $\mathcal{H} = \mathbb{E}[H]$ with $ H(X_1,X_2)=\int\limits_{X'\in\mathcal{R}}h(X_1,X')h^\ast(X_2,X')\mathrm{d}X'$. By definition, $H(X_1,X_2)$ is a Hermitian kernel with non-negative eigenvalues.

With this reformulation, one has an analytical solution for maximizing~\eqref{eqn:I} when constrained on~\eqref{eqn:I0}. As summarized in~\cite{schulz2005optimal}, when $\EE[H]$ is given, the optimal solution to~\eqref{eqn:optimization} has the form of a coherent mode, namely:
\[
J(X_1\,,X_2) = \alpha_1\psi_1(X_1)\psi_1^\ast(X_2)\,,\quad\text{with}\quad \alpha_1 = 1\,,
\]
where $\psi_1$ is the eigenfunction of $\EE[H]$ associated with the largest eigenvalue. Since finding eigenvalues and eigenfunctions of a given matrix is mathematically straightforward, the only obstacle of identifying the optimal beam configuration lies in formulating $\EE[H]$.

\section{Calculation of transfer function}
While many of the main ideas here are applicable to general types of waves, in what follows we consider here the setting of optics with the paraxial wave equation (PWE), for concreteness. The PWE is also known as the parabolic approximation
\cite{Tappert1977,radder_1979,white2010high}.

%\sam{Need to change section title to be, say, ``Calculation of $\calH$ in physical space''? Since Fourier space version, and $\calH$ properties, are now in a subsequent section?}

% \sam{Should we add some sentences to the beginning of each section, in order to orient the reader and remind the reader of the contents and goals of the section?}

% \sam{Question: Should we just set $n_r=1$ everywhere?}

As a starting point, we recall the form of the PWE:
\begin{equation}\label{eqn:PWE}
\nabla_\perp^2 A 
+2ik\frac{\partial A}{\partial z} + k^2\left(\frac{n^2}{n_r^2}-1\right)A
=0\,,
\end{equation}
where $A$ is the complex signal amplitude, $k$ is the reference wave number of the source, and $n=n(x,y,z)$ is the index of refraction, which is a function of turbulent anomalies of air temperature. Here $X=(x,y)$ denotes the perpendicular direction, with $\nabla_\perp^2=\partial_x^2+\partial_y^2$, and $z$ denotes the transverse direction of propagation. The source is placed at $z=0$ and the receiver is located at $z=Z$. As a consequence:
\[
A(X,z=0)=\phi(X)\,,\quad\text{and}\quad U(X') = A(X',z=Z)\,.
\]

%\sam{Should the latter be $U(X') = A(X',z=Z)$, with the prime in two places?}\an{fixed}
% \an{The PWE can been solved numerically to study light propagation, for example in~\cite{white2010high}\ql{this should not be here... should be in the numerical section} the authors used a high-order compact difference scheme to simulate beam propagation through turbulence.}
In the weak fluctuation regime, the refraction index of the medium centers around a constant $n_r$:
\begin{equation}
    n^2=n_r^2 + \epsilon n_1^2
    \label{eqn:n2-expand}
\end{equation}
where $\epsilon$ encodes the small amplitude of the fluctuation. \revision{(To arrive at (\ref{eqn:n2-expand}), start from the ansatz $n = n_r +\tilde{n}$, so that $n^2$ expands as $n^2 = n_r^2 + 2\tilde{n}n_r + \tilde{n}^2$. Under weak fluctuations, $\tilde{n}$ is substantially smaller than $n_r$. So we exclude $\tilde{n}^2$ from the calculations, and change notation from $2\tilde{n}n_r$ to $\epsilon n_1^2$.)} In this setting, we can use perturbation theory
\cite{andrews2005laser,clifford1978classical}, and \revision{follow the classical asymptotic expansion technique~\cite{orszag1978advanced,kevorkian2012multiple} to expand $A$ to be:}
\begin{equation}\label{eqn:expansion_A}
    A=A_0+\epsilon A_1+\epsilon^2 A_2+\cdots\,.
\end{equation}
Inserting this ansatz into~\eqref{eqn:PWE} and balancing each order of $\epsilon$, we have:
\begin{align}
\epsilon^0:&&
\nabla_\perp^2 A_0 
+2ik\frac{\partial A_0}{\partial z} 
&=0\,, \label{eqn:A0-pde}
\\
\epsilon^1:&&
\nabla_\perp^2 A_1 
+2ik\frac{\partial A_1}{\partial z} 
&=
-k^2\frac{n_1^2}{n_r^2}A_0\,,\label{eqn:A1-pde}
\\
\epsilon^2:&&
\nabla_\perp^2 A_2
+2ik\frac{\partial A_2}{\partial z} 
&=
-k^2\frac{n_1^2}{n_r^2}A_1\,.
\label{eqn:A2-pde}
\end{align}
Denoting $V(x,y,z)=\frac{n_1^2}{n_r^2}$, we can express the solutions to $A_i$ explicitly, as summarized in the following subsection.
%where we will use $V$ in place of $n_1^2/n_r^2$ to simplify notation.
\subsection{Hierarchical solution to PWE in the weak fluctuation regime}
%\sam{Below, should we define the Fourier transform and its inverse?}\an{ done}
Suppose $X\in\mathbb{R}^d$.
% We denote the Fourier transform of a function $f(\cdot, z)$ wrt $X$ by $\hat{f}(\cdot,z)$ (definition is in Appendix). 
% \sam{Also, should we move the definition of $V$ from above to below, to the place of its first use? Yes -- in fact, define $V$ way up where $n_1$ is first introduced. And did we decide to set $n_r=1$ everywhere? }
Recall the equation for $A_0$ in~\eqref{eqn:A0-pde} with $\phi$ as the initial source term:
\begin{equation}\label{eqn:A_0}
\nabla_\perp^2 A_0 
+2ik\frac{\partial A_0}{\partial z}=0\,,\quad A(X,0)=\phi(X)\,.
\end{equation}
For $K\in\mathbb{R}^d$, the Fourier transform is $\hat{f}(K,z)=\int\limits_{X\in\mathbb{R}^d}f(X,z)e^{-iK\cdot X}\mathrm{d}X$ and thus the inverse is $f(X,z)=\frac{1}{(2\pi)^d}\int\limits_{K\in\mathbb{R}^d}\hat{f}(K,z)e^{iK\cdot X}\mathrm{d}K$. We perform the Fourier transform of equation~\eqref{eqn:A_0} to have: $-\|K\|^2\hat{A}_0+2ik\frac{\partial\hat{A}_0}{\partial z}=0$. This ODE \revision{(ordinary differential equation)} has an explicit solution:
\begin{equation}\label{A0:sol_Fourier}
\hat{A}_0(K,z)=\hat{\phi}(K)\hat{G}(K,z)\,\quad \text{with}\quad
% \begin{equation}\label{eqn:G_hat}
    \hat{G}(K,z)=\exp(-\frac{iz}{2k}\|K\|^2)\,.
\end{equation}
$\hat{G}$ is the Fourier transform of the uniform-medium Green's function given by
\begin{equation}\label{eqn:Green's_function}
%   \begin{aligned}
    G(X,z)=\Big(\frac{1}{2\pi}\Big)^{d/2}\Big(\frac{k}{iz}\Big)^{d/2}\exp\Big(\frac{ik}{2z}\|X\|^2\Big)\,.
% \end{aligned}  
\end{equation}
Taking the inverse Fourier transform we obtain:
\begin{align}\label{A0:sol}
A_0(X,z)&=\phi*G(X,z)\,,
\end{align}
where $\ast$ is the convolution sign. Similarly, by applying the Fourier transform to the equation for $A_1$ in~\eqref{eqn:A1-pde}, and define $S(K,z)=\widehat{VA_0}(K,z)$, we have:
% Note that when $z=0$, we have $G$ as the inverse Fourier transform of 1, so $G(X,0)=\delta(X)$. 
% \begin{equation}\label{eqn:A1}
% \nabla_\perp^2 A_1 
% +2ik\frac{\partial A_1}{\partial z} =-k^2VA_0\,,\quad A_1(X,0)=0\,.
% \end{equation}
% and its Fourier transform satisfies
$-\|K\|^2\hat{A}_1+2ik\frac{\partial \hat{A}_1}{\partial z}=-k^2S$. Its solution is
\begin{align}\label{A1:sol_Fourier}
      & \hat{A}_1(K,z)=\frac{ik}{2}\int_{z'=0}^zS(K,z')\hat{G}(K,z-z')\mathrm{d}z'\,,
    %   \exp\big(-\frac{i\|K\|^2}{2k}(z-z')\big)\mathrm{d}z'\,,
 \end{align}
and consequently:
\begin{align}\label{A1:sol}
    A_1(X,z)&=\frac{ik}{2}\int_{z'=0}^z\left[(VA_0)(\cdot ,z')*G(\cdot ,z-z')\right](X)\mathrm{d}z'\,,
\end{align}
where $G(\cdot,z-z')$ denotes the Green's function evaluated at the shifted $z$ coordinate $z-z'$, and
\[
\left[VA_0(\cdot,z')\ast G(\cdot,z-z')\right](X) = \int_{X'\in\mathbb{R}^d}VA_0(X',z')G(X-X',z-z'))\mathrm{d}X'\,.
\]
The solution to $A_2$ is similarly obtained:
\begin{align}\label{A2:sol}
    A_2(X,z)&=\frac{ik}{2}\int_{z'=0}^zdz'\left[(VA_1)(\cdot ,z')*G(\cdot ,z-z')\right](X)\,.
\end{align}

\subsection{Calculation of the propagator}
From the definition in~\eqref{eqn:U} and the expansion in~\eqref{eqn:expansion_A}, we express
\begin{equation}\label{eqn:h_expansion}
h = h_0 + \epsilon h_1+\epsilon^2 h_2+\text{high orders}
\end{equation}
where $h_i(X,X')$ is the propagator that collects the contribution of $\phi(X)$ in $A_i(X',Z)$. According to~\eqref{A0:sol}, we have
\begin{equation}
U(X')=A_0(X',Z)=\int\limits_{X\in\mathcal{A}}\phi(X)G(X'-X,Z)\mathrm{d}X\,,
\end{equation}
which naturally makes
\begin{equation}\label{eqn:h0}
    h_0(X,X')=G(X'-X,Z)\,.
\end{equation}
For the next order, the rearrangement of~\eqref{A1:sol} gives
% \begin{equation*}
% \begin{aligned}
% A_1(X',Z)&=\frac{ik}{2}\int_{z_1=0}^Z\mathrm{d}z_1\int_{X_1'\in \mathbb{R}^2}(VA_0)(X_1',z_1)G(X'-X_1',Z-z_1)\mathrm{d}X_1'\\
% &=\frac{ik}{2}\int_{z_1=0}^Z\mathrm{d}z_1\int_{X_1'\in \mathbb{R}^2}V(X_1',z_1)G(X'-X_1',Z-z_1)\mathrm{d}X_1'\int\limits_{X\in\mathcal{A}}\phi(X)G(X_1'-X,z_1)\mathrm{d}X\,.
% \end{aligned}
% \end{equation*}
% This gives:
\begin{equation}\label{eqn:h1}
    h_1(X,X')=\frac{ik}{2}\int_{z_1=0}^Z\mathrm{d}z_1\int_{X_1'\in \mathbb{R}^d}V(X_1',z_1)G(X'-X_1',Z-z_1)G(X_1'-X,z_1)\mathrm{d}X_1'\,.
\end{equation}
One way to interpret this formula is to set a screen at $z=z_1$ and collect all plane wave contribution from $(z=0,X)$ to $(z_1,X'_1)$, and view them as wave fronts by continuing propagating the plane waves to $(z=Z,X')$, coinciding with the Huygens–Fresnel principle. This interpretation suggests that $h_1$ collects the information of waves that gets scattered once.

Similarly, we have the expression for $h_2$ as
\begin{equation}\label{eqn:h2}
\begin{aligned}
    h_2(X,X')&=-\frac{k^2}{4}\int\limits_{z_1=0}^Z\int\limits_{X_1'\in\mathbb{R}^d}V(X_1',z_1)G(X'-X_1',Z-z_1)\mathrm{d}X_1'\mathrm{d}z_1\\
    &\int\limits_{z_2=0}^{z_1}\int\limits_{X_2'\in\mathbb{R}^d}V(X_2',z_2)G(X_1'-X_2',z_1-z_2)G(X_2'-X,z_2)\mathrm{d}X_2'\mathrm{d}z_2
\end{aligned}
\end{equation}
Similar to the interpretation above, one can view it as a collection of waves that scatter twice, once at $(z_1,X_1')$ and once at $(z_2,X_2')$. The media information is included only through the scattering points.

\subsection{Calculation of the propagator on the Fourier space}
The calculation can be repeated on the Fourier domain as well. Since the presentation of equations~\eqref{eqn:A0-pde}-\eqref{eqn:A2-pde} on Fourier domain are all linear, there exists a propagator, termed $g$, so that the solution:
\begin{equation}\label{eqn:def_g}
\hat{U}(K)=\hat{A}(K,Z)=\int\limits_{K'\in\mathbb{R}^d}\hat{\phi}(K')g(K',K)\mathrm{d}K'\,.
\end{equation}
As done in the physical domain, we expand $g$ in terms of powers of $\epsilon$ to have
\begin{equation*}
    g = g_0 + \epsilon g_1 + \epsilon^2 g_2 + \text{high orders}\,,
\end{equation*}
with $g_i$ presenting the contribution from $\hat{A}_i$. Comparing to~\eqref{A0:sol_Fourier} it is straightforward to have:
\begin{equation}\label{eqn:g_0}
    g_0(K',K)=\hat{G}(K,Z)\delta(K-K')\,.
\end{equation}

To compute $g_{1}$, we recall~\eqref{A1:sol_Fourier} and plug in~\eqref{A0:sol_Fourier}:
\begin{equation}
    \begin{aligned}
    \hat{U}_1(K)&=\frac{ik}{2}\int\limits_{z_1=0}^ZS(K,z_1)\hat{G}(K,Z-z_1)\mathrm{d}z_1\\
    &=\frac{ik}{2(2\pi)^d}\int\limits_{z_1=0}^Z\hat{G}(K,Z-z_1)\mathrm{d}z_1\int\limits_{K'\in\mathbb{R}^d}\hat{V}(K-K',z_1)\hat{G}(K',z_1)\hat{\phi}(K')\mathrm{d}K'\,,
    \end{aligned}
\end{equation}
which suggests:
\begin{equation}\label{eqn:g_1}
    g_1(K',K)=\frac{ik}{2(2\pi)^d}\int\limits_{z_1=0}^Z\hat{G}(K,Z-z_1)\hat{G}(K',z_1)\hat{V}(K-K',z_1)\mathrm{d}z_1\,.
\end{equation}

Similarly, we have
\begin{equation}\label{eqn:g_2}
\begin{aligned}
    g_2(K',K)=&-\Big(\frac{1}{2\pi}\Big)^{2d}\frac{k^2}{4}\int\limits_{z_1=0}^Z\hat{G}(K,Z-z_1)\mathrm{d}z_1\int\limits_{K''\in\mathbb{R}^d}\hat{V}(K-K'',z_1)\mathrm{d}K''\\
    &\times\int\limits_{z_2=0}^{z_1}\hat{G}(K'',z_1-z_2)\hat{G}(K',z_2)\hat{V}(K''-K',z_2)\mathrm{d}z_2\,.
\end{aligned}
\end{equation}
The formula~\eqref{eqn:g_0},~\eqref{eqn:g_1} and~\eqref{eqn:g_2} are suggesting that the scattering takes place at location $z_1$ for once-scattered wave, and $z_1$, $z_2$ for twice-scattered wave respectively, and only through scattering, the wave picks up the frequency information from the media.

\section{Calculation of $\calH$}
The explicit formulation of the propagators, $h$ on the physical space and $g$ on the Fourier domain, allows us to compute $\mathcal{H}$. We derive the formula in this section.
\subsection{Representation in physical space }\label{sec:calH_physical}
Noting the definition~\eqref{eqn:calH} for the expansion in~\eqref{eqn:h_expansion}, we rewrite:
\begin{equation}
\calH=\calH_{00}+\epsilon^2\big(\calH_{02}+\calH_{11}+\calH_{20}\big)\,,
\end{equation}
where $\calH_{ij}(X_1,X_2)=\mathbb{E}\Big[\int\limits_{X'\in\mathcal{R}}h_i(X_1,X')h^\ast_j(X_2,X')\mathrm{d}X'\Big]$. We should note that $\calH_{01}=0$ and $\calH_{10}=0$ and thus are dropped out of the expansion. We now calculate each term. First, recalling~\eqref{eqn:h0}, we have:
\begin{equation}\label{eqn:H00}
\begin{aligned}
\calH_{00}(X_1,X_2)&=\int\limits_{X'\in\mathcal{R}}G(X_1-X',Z)G^\ast(X_2-X',Z)\mathrm{d}X'\\
&=\Big(\frac{1}{2\pi}\Big)^d\Big(\frac{k}{Z}\Big)^d\int\limits_{X'\in\mathcal{R}}\exp\Big(\frac{ik}{2Z}(\|X_1-X'\|^2-\|X_2-X'\|^2)\Big)\mathrm{d}X'\\
&=G(X_1,Z)G^\ast(X_2,Z)\int\limits_{X'\in\mathcal{R}}\exp\Big(-\frac{ik}{Z}X'\cdot(X_1-X_2)\Big)\mathrm{d}X'\,.
\end{aligned}
\end{equation}
Similarly, defining
\begin{equation}\label{eqn:Gamma}
\mathbb{E}[V(X_1,z_1)V^\ast(X_2,z_2)]=\Gamma_V(X_1,z_1,X_2,z_2)\,,
\end{equation}
and plugging~\eqref{eqn:h1} in $\mathcal{H}_{11}$, we have:
\begin{equation}\label{eqn:H11}
\begin{aligned}
\calH_{11}(X_1,X_2)
% &=\frac{k^2}{4}\int\limits_{z_1,z_2=0}^Z\mathrm{d}z_1\mathrm{d}z_2\int\limits_{\mathcal{R}^d\times\mathcal{R}^d}\mathbb{E}[V(X_1',z_1)V^\ast(X_2',z_2)]G(X_1-X_1',z_1)G^\ast(X_2-X_2',z_2)\mathrm{d}X_1'\mathrm{d}X_2'\\
% &\times\int\limits_{X'\in\mathcal{R}}G(X'-X_1',Z-z_1)G^\ast(X'-X_2',Z-z_2)\mathrm{d}X'\\
&=\frac{k^2}{4}\int\limits_{z_1,z_2=0}^Z\mathrm{d}z_1\mathrm{d}z_2\int\limits_{\mathbb{R}^d\times \mathbb{R}^d}\Gamma_V(X_1',z_1,X_2',z_2)G(X_1-X_1',z_1)G^\ast(X_2-X_2',z_2)\mathrm{d}X_1'\mathrm{d}X_2'\\
&\times\int\limits_{X'\in\mathcal{R}}G(X'-X_1',Z-z_1)G^\ast(X'-X_2',Z-z_2)\mathrm{d}X'\,,
\end{aligned}
\end{equation}
% where we set the notation . 
% Then we have
% \begin{equation}\label{eqn:H11}
%     \begin{aligned}
%             \calH_{11}(X_1,X_2)&=\frac{k^2}{4}\int\limits_{z_1=0}^Z\int\limits_{z_2=0}^Z\mathrm{d}z_1\mathrm{d}z_2\int\limits_{X_1'\in\mathcal{R}^d}\int\limits_{X_2'\in\mathcal{R}^d}\Gamma_V(X_1',z_1,X_2',z_2)G(X_1-X_1',z_1)G^\ast(X_2-X_2',z_2)\mathrm{d}X_1'\mathrm{d}X_2'\\
%             &\times\int\limits_{X'\in\mathcal{R}}G(X'-X_1',z_1)G^\ast(X'-X_2',z_2)\mathrm{d}X'
%     \end{aligned}
% \end{equation}
The formulations of $\calH_{02}$ and $\calH_{20}$ naturally follow as
% Similarly we can write down the rest of the expressions as
% \begin{equation}
%     \begin{aligned}
%             \calH_{02}(X_1,X_2)&=-\frac{k^2}{4}\int\limits_{z_1=0}^Z\int\limits_{z_2=0}^{z_1}\mathrm{d}z_2\mathrm{d}z_1\int\limits_{X_1'\in\mathbb{R}^d}\int\limits_{X_2'\in\mathbb{R}^d}\mathbb{E}[V^\ast(X_1',z_1)V^\ast(X_2',z_2)]G^\ast(X_1'-X_2',z_1-z_2)G^\ast(X_2'-X_2,z_2)\mathrm{d}X_1'\mathrm{d}X_2'\\
%             &\times\int\limits_{X'\in\mathcal{R}}G(X'-X_1,Z)G^\ast(X'-X_1',Z-z_1)\mathrm{d}X'
%     \end{aligned}
% \end{equation}
% Substituting the covariance function in the above expression gives
\begin{equation}\label{eqn:H02}
\begin{aligned}
\calH_{02}(X_1,X_2)&=-\frac{k^2}{4}\int\limits_{z_1=0}^Z\int\limits_{z_2=0}^{z_1}\mathrm{d}z_2\mathrm{d}z_1\int\limits_{X_1'\in\mathbb{R}^d}\int\limits_{X_2'\in\mathbb{R}^d}\Gamma_V(X_1,'z_1,X_2',z_2)G^\ast(X_1'-X_2',z_1-z_2)G^\ast(X_2'-X_2,z_2)\mathrm{d}X_1'\mathrm{d}X_2'\\
&\times\int\limits_{X'\in\mathcal{R}}G(X'-X_1,Z)G^\ast(X'-X_1',Z-z_1)\mathrm{d}X'\,,
\end{aligned}
\end{equation}
and $\calH_{20}=\calH^\ast_{02}$. We note that this is a very complicated formulation and can hardly be of practical use in reality. More specifically, for each fixed $(X_1,X_2)$, the computation of $\calH_{02}$, for example, amounts to $8$ dimensional integral when $d=2$.
% For each configuration of $(X_1,X_2)\in\mathcal{A}^2$, we need to calculate all $\calH_{ij}$, with each term being a five-fold integral. If $d=2$, the integration is conducted in an $8$-dimensional space. Computationally this makes the problem infeasible.

\subsection{Representation in the Fourier domain}\label{sec:calH_Fourier}
% It is immediate that the computation of $\calH_{ij}$ is computationally infeasible. Each term is a five-fold integral. For $d=2$, this leads to an integration on an $8$-dimensional space.

The computational cost is prohibitive if the calculation is conducted on the physical domain. However, in reality, assumptions on the atmospheric turbulence are typically represented on the Fourier domain of $\Gamma_V$. Naturally, if one can repeat the process on the Fourier space and incorporate the assumptions, computational difficulty can potentially be reduced. We explore such possibility in this section.

% The technical difficulty is immediate when the propagator is calculated in physical space, as described above. However, certain assumptions on atmospheric turbulence can simplify the Fourier transform of $\Gamma_V$; therefore, to fully use these assumptions, our computation should be conducted on the Fourier domain as well.

%\sam{Below, in our definition of $I_\mathcal{R}$, are we missing $\langle \cdot\rangle$ for averaging over the source fluctuations? Need this in order to connect $\phi$ and $J$ later on below?}\an{ done. $J$ is defined in terms of $\phi$ at a later step.}

Recall that the intensity to be maximized is:
\begin{equation*}
I_\mathcal{R}=\int\limits_{X'\in\mathcal{R}}\EE\left[\langle U(X')U^\ast(X')\rangle\right]\mathrm{d}X'=\int\limits_{X'\in\mathbb{R}^d}\EE\left[\langle U(X')w(X')U^\ast(X')w^\ast(X')\rangle\right]\mathrm{d}X'\,,
\end{equation*}
where $w$ is the indicator function that takes value $1$ within the window $x\in\mathcal{R}$, and $0$ outside. Using Parseval's equality, this translates to:
\begin{equation}\label{eqn:I_W}
I_\mathcal{R}=\frac{1}{(2\pi)^d}\int\limits_{K\in\mathbb{R}^d}\EE\left[\langle W(K)W^\ast(K)\rangle\right]\mathrm{d}K\,,
\end{equation}
where $W=\widehat{Uw}$ denotes the Fourier transform:
\begin{equation}
    W(K)=\widehat{Uw}(K) = \frac{1}{(2\pi)^d}\hat{U}\ast\hat{w}\, ,
\end{equation}
Expand this convolution in~\eqref{eqn:I_W}:
\begin{equation}
% \begin{aligned}
I_\mathcal{R}
% &=\frac{1}{(2\pi)^{3d}}\int\limits_{K\in\mathbb{R}^d}\mathbb{E}\Big[\langle\big(\hat{U}\ast\hat{w}\big)(K)\big(\hat{U}^\ast\ast\hat{w}^\ast\big)(K)\rangle\Big]\mathrm{d}K\\
    =\frac{1}{(2\pi)^{3d}}\int\limits_{K_1,K_2\in\mathbb{R}^d}\mathbb{E}\big[\langle\hat{U}(K_1)\hat{U}^\ast(K_2)\rangle\big]\left( \;\int\limits_{K\in\mathbb{R}^d}\hat{w}(K-K_1)\hat{w}^\ast(K-K_2)\mathrm{d}K \right) \mathrm{d}K_1\mathrm{d}K_2\,.
% \end{aligned}
\end{equation}
We note that the window function term can be simplified:
\begin{equation*}
    \begin{aligned}
   \frac{1}{(2\pi)^{d}} \int\limits_{K\in\mathbb{R}^d}\hat{w}(K-K_1)\hat{w}^\ast(K-K_2)\mathrm{d}K&=\int\limits_{X'\in\mathcal{R}}\int\limits_{X''\in\mathcal{R}}\mathrm{d}X'\mathrm{d}X''e^{i(K_1\cdot X'-K_2\cdot X'')}\delta(X'-X'')\\
    &=\int\limits_{X'\in\mathcal{R}}\mathrm{d}X'e^{i(K_1-K_2)\cdot X'}=\hat{w}(K_1-K_2)\,.
%   \frac{1}{(2\pi)^{d}} \int\limits_{K\in\mathbb{R}^d}\hat{w}(K-K_1)\hat{w}^\ast(K-K_2)\mathrm{d}K&=\frac{1}{(2\pi)^{d}}\int\limits_{K\in\mathbb{R}^d}\mathrm{d}K\int\limits_{X'\in\mathcal{R}}\int\limits_{X''\in\mathcal{R}}\mathrm{d}X'\mathrm{d}X''e^{-i(K-K_1)\cdot X'}e^{i(K-K_2)\cdot X''}\\
%     % &=\int\limits_{X'\in\mathcal{R}}\int\limits_{X''\in\mathcal{R}}\mathrm{d}X'\mathrm{d}X''e^{i(K_1\cdot X'-K_2\cdot X'')}\int\limits_{K\in\mathbb{R}^d}\mathrm{d}Ke^{-i K\cdot(X'-X'')}\\
%     &=\int\limits_{X'\in\mathcal{R}}\int\limits_{X''\in\mathcal{R}}\mathrm{d}X'\mathrm{d}X''e^{i(K_1\cdot X'-K_2\cdot X'')}\delta(X'-X'')\\
%     &=\int\limits_{X'\in\mathcal{R}}\mathrm{d}X'e^{i(K_1-K_2)\cdot X'}=w_M(K_1-K_2)\,.
    \end{aligned}
\end{equation*}
%\sam{Why did we start to use the notation $w_M$?  Didn't we call it $\hat{w}$ a few equations ago? These are both the Fourier transform of the indicator function of the receiver region, right?}\an{ I think that is true when the reciever region is symmetric. changed notation to $\hat{w}$}
and that $\hat{U}(K)=\sum_j\epsilon^j\hat{U}_j(K)$, we adopt the representation~\eqref{eqn:def_g} to further rewrite $I_\mathcal{R}$ to:
\begin{equation}\label{eqn:I_R_Fourier1}
\begin{aligned}
I_\mathcal{R}
% &= \frac{1}{(2\pi)^{2d}}\int\limits_{K_1\in\mathbb{R}^d}\int\limits_{K_2\in\mathbb{R}^d}\mathbb{E}\big[\langle\hat{U}(K_1)\hat{U}^\ast(K_2)\rangle\big]\hat{w}(K_1-K_2)\mathrm{d}K_1\mathrm{d}K_2\\
&=\sum_{ij}\frac{\epsilon^{i+j}}{(2\pi)^{2d}}\int\limits_{K_1,K_2\in\mathbb{R}^d}\mathbb{E}\big[\langle\hat{U}_i(K_1)\hat{U}_j^\ast(K_2)\rangle\big]\hat{w}(K_1-K_2)\mathrm{d}K_1\mathrm{d}K_2\\
% \end{aligned}
% \end{equation*}
% Using representation~\eqref{eqn:def_g}, we rewrite $I_\mathcal{R}$ as
% % Interchanging $K_1, K_2$ with $K_1', K_2'$ respectively, we have
% \begin{equation}\label{eqn:I_R_Fourier1}
%     \begin{aligned}
    % I_\mathcal{R}
    &=\sum_{ij}\frac{\epsilon^{i+j}}{(2\pi)^{2d}}\int\limits_{K_1,K_2\in\mathbb{R}^d}\langle\hat{\phi}(K_1)\hat{\phi}^\ast(K_2)\rangle\mathrm{d}K_1\mathrm{d}K_2\\
&\times\int\limits_{K_1',K_2'\in\mathbb{R}^d}\mathbb{E}\big[g_i(K_1,K_1')g_j^\ast(K_2,K_2')\big]\hat{w}(K_1'-K_2')\mathrm{d}K_1'\mathrm{d}K_2'\,.
    \end{aligned}
\end{equation}
Recall that $J(X_1,X_2)=\langle \phi(X_1)\phi^*(X_2)\rangle$ and defining
\begin{equation}
    \hat{J}(K_1,K_2)=\int\limits_{X_1\in\mathbb{R}^d}\int\limits_{X_2\in\mathbb{R}^d}J(X_1,X_2)e^{-i(K_1\cdot X_1-K_2\cdot X_2)}\mathrm{d}X_2\mathrm{d}X_1\,,
\end{equation}
equation~\eqref{eqn:I_R_Fourier1} becomes
\begin{equation}\label{eqn:I_R_hat}
    \begin{aligned}
    I_\mathcal{R}=&\sum_{ij}\frac{\epsilon^{i+j}}{(2\pi)^{2d}}\int\limits_{K_1,K_2\in\mathbb{R}^d}\hat{J}(K_1,K_2)\mathrm{d}K_1\mathrm{d}K_2\\
&\times\int\limits_{K_1',K_2'\in\mathbb{R}^d}\mathbb{E}\big[g_i(K_1,K_1')g_j^\ast(K_2,K_2')\big]\hat{w}(K_1'-K_2')\mathrm{d}K_1'\mathrm{d}K_2'\\
=&\frac{1}{(2\pi)^d}\int\limits_{K_1,K_2\in\mathcal{R}^d}\hat{J}(K_1,K_2)\calH_{\rmf}(K_1,K_2)\mathrm{d}K_1\mathrm{d}K_2\,,
    \end{aligned}
\end{equation}
where we define
\begin{equation}\label{eqn:def_hatH}
\begin{aligned}
\calH_{\rmf,ij}&=\frac{1}{(2\pi)^d}\int\limits_{K_1'\in\mathbb{R}^d}\int\limits_{K_2'\in\mathbb{R}^d}\mathbb{E}\big[g_i(K_1,K_1')g_j^\ast(K_2,K_2')\big]\hat{w}(K_1'-K_2')\mathrm{d}K_1'\mathrm{d}K_2'\,,
\end{aligned}
\end{equation}
and set
\[
\calH_{\rmf}=\calH_{\rmf,00}+\epsilon^2\Big(\calH_{\rmf,02}+\calH_{\rmf,11}+\calH_{\rmf,20}\Big)\,.
\]

With the same argument as provided in~\cite{schulz2005optimal}, to produce the optimal beam that provides the maximum light intensity, $\hat{J}$ should be composed of coherent beam, in the sense that
\[
\hat{J}(K_1,K_2)=\psi(K_1)\psi^\ast(K_2)\,,
\]
where $\psi(K)$ is the eigenfunction of $\calH_{\rmf}$ associated with the largest eigenvalue. The statement translates the problem to computing $\calH_{\rmf}$ and its eigenfunctions. This can be readily done. Recall~\eqref{eqn:g_0} we have:
\begin{equation}\label{eqn:H00_Fourier}
\begin{aligned}
\calH_{\rmf,00}(K_1,K_2)&=\frac{1}{(2\pi)^d}\hat{G}(K_1,Z)\hat{G}^\ast(K_2,Z)\hat{w}(K_1-K_2)\,.
\end{aligned}
\end{equation}
To find $\calH_{\rmf,11}$, we use~\eqref{eqn:g_1}:
\begin{equation}\label{eqn:H11_Fourier}
    \begin{aligned}
     &\calH_{\rmf,11}(K_1,K_2)=\Big(\frac{1}{2\pi}\Big)^{3d}\frac{k^2}{4}\int_{z_1=0}^Z\int_{z_2=0}^Z\hat{G}(K_1,z_1)\hat{G}^\ast(K_2,z_2)\mathrm{d}z_2\mathrm{d}z_1\\
     &\times\int\limits_{K_1',K_2'\in\mathbb{R}^d}\hat{G}(K_1',Z-z_1)\hat{G}^\ast(K_2',Z-z_2)\hat{\Gamma}_V(K_1'-K_1,z_1,K_2'-K_2,z_2)\hat{w}(K_1'-K_2')\mathrm{d}K_1'\mathrm{d}K_2'
    \end{aligned}
\end{equation}
% \begin{equation}
%     \begin{aligned}
%     \mathbb{E}[g_1(K_1',K_1)g_2(K_2',K_2)]&=\Big(\frac{1}{2\pi}\Big)^{2d}\frac{k^2}{4}\int_{z_1=0}^Z\int_{z_2=0}^Z\hat{G}(K_1',Z-z_1)\hat{G}^\ast(K_2',Z-z_2)\\
%     &\times G(K_1,z_1)G^\ast(K_2,z_2)\hat{\Gamma}_V(K_1'-K_1,z_1,K_2'-K_2,z_2)\mathrm{d}z_2\mathrm{d}z_1\,.
%     \end{aligned}
% \end{equation}
Here $\hat{\Gamma}_V$ is the covariance of the random medium in Fourier space, given by
\begin{equation}
  \hat{\Gamma}_V(K_1,z_1,K_2,z_2)=\mathbb{E}\big[\hat{V}(K_1,z_1)\hat{V}^\ast(K_2,z_2)\big]\,.
\end{equation}

Similarly, using~\eqref{eqn:g_2} we have
% \begin{equation}
%     \begin{aligned}
%     &\mathbb{E}[g_0(K_1',K_1)g^\ast_2(K_2',K_2)]=-\Big(\frac{1}{2\pi}\Big)^{2d}\frac{k^2}{4}\hat{G}(K_1',Z)\delta(K_1'-K_1)\int\limits_{z_1=0}^Z\hat{G}(K_2',Z-z_1)\mathrm{d}z_1\\
%     &\times\int\limits_{z_2=0}^Z\int\limits_{K''\in\mathbb{R}^d}\hat{G}(K'',z_1-z_2)\hat{G}(K_2,z_2)\hat{\Gamma}_V(K_2'-K'',z_1,K_2-K'',z_2)\mathrm{d}K''\mathrm{d}z_2
%     \end{aligned}
% \end{equation}
% and rearranging the dummy indices of integration we have
\begin{equation}\label{eqn:H02_Fourier}
    \begin{aligned}
    &\calH_{\rmf,02}(K_1,K_2)=-\Big(\frac{1}{2\pi}\Big)^{3d}\frac{k^2}{8}\hat{G}(K_1,Z)\int\limits_{z_1=0}^Z\int\limits_{z_2=0}^{Z}\hat{G}^\ast(K_2,z_2)\mathrm{d}z_2\mathrm{d}z_1\\
    &\times\int\limits_{K_1',K_2'\in\mathbb{R}^d}\hat{G}^\ast(K_2',z_1-z_2)\hat{G}^\ast(K_1',Z-z_1)\hat{\Gamma}_V(K_2-K_2',z_2,K_1'-K_2',z_1)\hat{w}(K_1-K_1')\mathrm{d}K_1'\mathrm{d}K_2'\\
    \end{aligned}
    \end{equation}
    Similarly
    \begin{equation}\label{eqn:H20_Fourier}
    \begin{aligned}
    &\calH_{\rmf,20}(K_1,K_2)=\calH_{\rmf,02}^\ast(K_1,K_2).
    \end{aligned}
    \end{equation}
    
    % \begin{equation}\label{eqn:H20_Fourier}
    % \begin{aligned}
    % &\calH_{\rmf,20}(K_1,K_2)=-\Big(\frac{1}{2\pi}\Big)^{3d}\frac{k^2}{4}\hat{G}^\ast(K_2,Z)\int\limits_{z_2=0}^Z\int\limits_{z_1=0}^{z_2}\hat{G}(K_1,z_2)\mathrm{d}z_1\mathrm{d}z_2\\
    % &\times\int\limits_{K_1',K_2'\in\mathbb{R}^d}\hat{G}(K_2',z_1-z_2)\hat{G}(K_1',Z-z_1)\hat{\Gamma}_V(K_1'-K_2',z_1,K_1-K_2',z_2)\hat{w}(K_1'-K_2)\mathrm{d}K_1'\mathrm{d}K_2'\,.
    % \end{aligned}
    % \end{equation}
We end our discussion by pointing out that even before any assumptions on the turbulence gets incorporated, the Fourier domain quantity $\calH_{\rmf,ij}$ is already easier than that of the physical-space quantity $\calH_{ij}$. Indeed, as presented in~\eqref{eqn:H11_Fourier}-\eqref{eqn:H20_Fourier}, the $\calH_{\rmf,ij}$ quantities are four-folded integrals once $\hat{\Gamma}$ is given. The extra integral in the $\calH_{ij}$ formula in~\eqref{eqn:H02} in $X\in\mathcal{R}$ is absorbed in $\hat{w}$ and has been completed analytically. More specifically, if $d=2$, for every fixed $(K_1,K_2)$, the calculation of $\calH_{\rmf,ij}$ is a $6$-dimensional integration, as compared to $8$-dimensional as shown in~\eqref{eqn:H02}.

\section{Properties of $\calH$}\label{sec:properties}

\subsection{Relation between $\calH$ and $\calH_{\rmf}$}
While $\calH$ and $\calH_{\rmf}$ were computed above from different perspectives (in physical space and Fourier space), they are related. To see the connection between them, note that the total average intensity can be written in terms of the mutual intensity function as
\begin{equation}
    \begin{aligned}
         I&=\frac{1}{(2\pi)^d}\int\limits_{K_1\in\mathbb{R}^d}\int\limits_{K_2\in\mathbb{R}^d}\hat{J}(K_1,K_2)\calH_{\rmf}(K_1,K_2)\mathrm{d}K_1\mathrm{d}K_2\\
         &=\frac{1}{(2\pi)^d}\int\limits_{K_1\in\mathbb{R}^d}\int\limits_{K_2\in\mathbb{R}^d}\calH_{\rmf}(K_1,K_2)\mathrm{d}K_1\mathrm{d}K_2\int\limits_{X_1\in\mathcal{A}}\int\limits_{X_2\in\mathcal{A}}J(X_1,X_2)e^{-i(K_1\cdot X_1-K_2\cdot X_2)}\mathrm{d}X_1\mathrm{d}X_2\\
         &=\frac{1}{(2\pi)^d}\int\limits_{X_1\in\mathcal{A}}\int\limits_{X_2\in\mathcal{A}}J(X_1,X_2)\mathrm{d}X_1\mathrm{d}X_2\int\limits_{K_1\in\mathbb{R}^d}\int\limits_{K_2\in\mathbb{R}^d}\calH_{\rmf}(K_1,K_2)e^{-i(K_1\cdot X_1-K_2\cdot X_2)}\mathrm{d}K_1\mathrm{d}K_2\\
    \end{aligned}
\end{equation}
which shows that
\begin{equation}\label{eqn:relation_H_f_p}
\calH(X_1,X_2)=\frac{1}{(2\pi)^d}\int\limits_{K_1\in\mathbb{R}^d}\int\limits_{K_2\in\mathbb{R}^d}\calH_{\rmf}(K_1,K_2)e^{-i(K_1\cdot X_1-K_2\cdot X_2)}\mathrm{d}K_1\mathrm{d}K_2\, ,
\end{equation}
so that $\calH$ and $\calH_{\rmf}$ are themselves related by a 
Fourier transformation.

\subsection{Energy conservation}

We now show that, by keeping terms up to $O(\epsilon^2)$ in 
the expansion, energy is conserved. This is an important reason
for keeping the $O(\epsilon^2)$ terms in addition to the 
$O(\epsilon)$ terms. Indeed,
one criticism over employing perturbation theory in wave propagation is that it often loses preservation of some crucial physical properties, such as energy conservation~\cite{charnotskii2015extended}. 
%This is a valid concern: The higher orders are regarded negligible and eliminated from the computation, and thus in the regime where the asymptotic expansion is no longer accurate, the physical quantities can be far away from the ground-truth. However, the formulation that we derive preserves the energy. 
However, energy conservation is retained here.

To see so, let the receiver occupy the entire space (i.e., $\mathcal{R}=\mathbb{R}^d$), 
in which case the window function becomes:
\begin{equation}
\hat{w}(K_1-K_2)=\int\limits_{X'\in\mathcal{R}^d}\mathrm{d}X'e^{i(K_1-K_2)\cdot X'}=(2\pi)^{d}\delta(K_1-K_2)\,.
\end{equation}
Inserting this into~\eqref{eqn:H00_Fourier} and \eqref{eqn:H11_Fourier}-\eqref{eqn:H20_Fourier}, we have
\begin{equation}
    \calH_{\rmf,00}(K_1,K_2)=\delta(K_1-K_2)\,,\quad \text{and}\quad \calH_{\rmf,11}+\calH_{\rmf,02}+\calH_{\rmf,20}=0\,,
\end{equation}
% and
% \begin{equation}
%     \calH_{\rmf,11}+\calH_{\rmf,02}+\calH_{\rmf,20}=0
% \end{equation}
so that
\begin{equation}
\begin{aligned}
I&=\frac{1}{(2\pi)^d}\int\limits_{K_1\in\mathbb{R}^d}\int\limits_{K_1\in\mathbb{R}^d}\hat{J}(K_1,K_2)\delta(K_1-K_2)\mathrm{d}K_1\mathrm{d}K_2\\
&=\frac{1}{(2\pi)^d}\int\limits_{K_1\in\mathbb{R}^d}\hat{J}(K_1,K_1)\mathrm{d}K_1=\int\limits_{X\in\mathcal{A}}J(X,X)\mathrm{d}X=I_0\,.
\end{aligned}
\end{equation}
We note that the energy conservation holds independent of $\epsilon$. This means that, even when $\epsilon$ and the associated approximation errors are relatively large, energy conservation still holds true and provides a certain level of physical realism.

\section{Some {optional} simplifications}\label{sec:assumptions}
One advantage of utilizing the Fourier-space presentation is that it makes it much easier to incorporate the classical assumptions on atmosphere turbulence since these assumptions are typically specified in the Fourier space. This would allow us to further reduce the computation.
% When the media randomness presents special structures, there is hope that one can reduce the computation even further.

%\sam{Would this be a good place to clarify which of the assumptions will be used in our numerical experiments? Do we use all of the assumptions below?  In our Fourier-space calculations, but not in our physical-space calculations? Maybe we can add a statement as follows:}

In this subsection, we describe several common assumptions and the resulting simplifications to the formulas for the $\calH_{\rmf,ij}$ quantities. Then, in a later section, a selection of these cases will be investigated via numerical calculations.

\subsection{Homogeneous-statistics assumption}\label{sec:homog_assumption}
One classical assumption is the homogeneous (or stationary) property.
% on the $X$-plane. 
This is to assume the covariance $\Gamma_V$ of the medium has the structure of
\begin{equation}\label{eqn:assume_1}
\Gamma_V(X_1,z_1,X_2,z_2)=f(X_1-X_2,z_1-z_2)\, .
\end{equation}
In this case, we also have
\begin{equation}
\hat{\Gamma}_V(K_1,z_1,K_2,z_2)=(2\pi)^dF(K_1,z_1-z_2)\delta(K_1-K_2)\,,
\end{equation}
where $F(K,z)=\hat{f}(X,z)$ is the Fourier transform. This newly induced $\delta$ function in the $K$ domain helps to eliminate one-fold of integration. For example, when inserted  into~\eqref{eqn:H11_Fourier}, we have:
\begin{equation}\label{eqn:H11_Fourier_new}
    \begin{aligned}
         \calH_{\rmf,11}(K_1,K_2)&=\frac{k^2}{4(2\pi)^{d}}\calH_{\rmf,00}(K_1,K_2)\int_{z_1=0}^Z\int_{z_2=0}^Z\mathrm{d}z_2\mathrm{d}z_1\\
     &\times\int\limits_{K'\in\mathbb{R}^d}F(K',z_1-z_2)\hat{G}(K',z_1-z_2)\exp\Big(-\frac{i}{k}K'\cdot(K_1z_1-K_2z_2)\Big)\mathrm{d}K'\\
     &=\frac{k^2}{4(2\pi)^{d}}\calH_{\rmf,00}(K_1,K_2)\int_{z_1=0}^Z\int_{z_2=0}^Z\widehat{F\hat{G}}\big((K_1z_1-K_2z_2)/k,z_1-z_2\big)\mathrm{d}z_2\mathrm{d}z_1\,.
     \end{aligned}
\end{equation}
We rewrote the innermost integral in terms of a Fourier transform in the last equation.

Similarly,
\begin{equation}\label{eqn:H02_Fourier_new}
    \begin{aligned}
    \calH_{\rmf,02}(K_1,K_2)&=-\frac{k^2}{8(2\pi)^{d}}\calH_{\rmf,00}(K_1,K_2)\int_{z_1=0}^Z\int_{z_2=0}^{Z}\widehat{F\hat{G}}\big(K_2(z_2-z_1)/k,z_1-z_2\big)\mathrm{d}z_2\mathrm{d}z_1\,.
\end{aligned}
\end{equation}
and $\calH_{\rmf,20}(K_1,K_2)=\calH_{\rmf,02}^\ast(K_1,K_2)$. We note that this has further simplified the computation to a $4$-dimensional integral, with two dimensions absorbed into the Fourier transform when $d=2$. Moreover, the Fourier transform component, $\widehat{F\hat{G}}$, though being a $2$-dimensional integral, can be calculated through FFT, which further reduces an $N^2$ computational cost to $N\log{N}$.

% \vspace{12pt}
% \an{added later} Equivalently, the innermost integrals can be written in terms of Fourier transforms so that
% \begin{equation}\label{eqn:Hf_Fourier_transform_simplification}
%     \begin{aligned}
%         \calH_{f11}(K_1,K_2)&=\frac{k^2}{4(2\pi)^{d}}\calH_{\rmf,00}(K_1,K_2)\int_{z_1=0}^Z\int_{z_2=0}^Z\widehat{F\hat{G}}\big((K_1z_1-K_2z_2)/k,z_1-z_2\big)\mathrm{d}z_2\mathrm{d}z_1\\
%         \calH_{f02}(K_1,K_2)&=-\frac{k^2}{4(2\pi)^{d}}\calH_{\rmf,00}(K_1,K_2)\int_{z_1=0}^Z\int_{z_2=0}^{z_1}\widehat{F\hat{G}}\big(K_2(z_2-z_1)/k,z_1-z_2\big)\mathrm{d}z_2\mathrm{d}z_1\\
%         \calH_{\rmf,20}(K_1,K_2)&=-\frac{k^2}{4(2\pi)^{d}}\calH_{\rmf,00}(K_1,K_2)\int_{z_2=0}^{Z}\int_{z_1=0}^{z_2}\widehat{F\hat{G}^\ast}\big(K_1(z_2-z_1)/k,z_1-z_2\big)\mathrm{d}z_1\mathrm{d}z_2
%     \end{aligned}
% \end{equation}
% \sam{Why were these equations added later? Are they necessary?} \an{I did use this formulation for numerical computation}

%\subsubsection{??\ql{can we call it the use of Kolmogorov spectrum?}}

\subsection{Decorrelation-in-$z$ assumption}

Another assumption that one might choose to make, in addition to the earlier assumption of homogeneity, is regarding the decay rate of the covariance in $z$. This assumption gets widely used, for instance, in~\cite{clifford1978classical}. It states that for some characteristic length $\delta_z\ll Z$,
\begin{equation}\label{eqn:assume_2}
F(K,z)\sim 0\,,\quad \mbox{for}\quad |z|>\delta_z\,.
\end{equation}
Note that, for comparison, this is a slightly relaxed assumption compared to the Markov approximation. If the Markovian approximation is assumed in the $z$ direction, then the random medium at every $z$ point is statistically independent. The assumption in (\ref{eqn:assume_2}), in contrast, allows a non-trivial correlation length in the $z$ direction, up to the length scale $\delta_z$.

To utilize this additional assumption, it is helpful to 
define the transformation
\begin{equation}
    \eta=\frac{z_1+z_2}{2}, \quad \mu=z_1-z_2.
\end{equation}
We then perform the change of variable to have
\begin{equation}\label{eqn:H_f_11_dec_z}
\begin{aligned}
\calH_{\rmf,11}(K_1,K_2)=&\frac{k^2}{4(2\pi)^{d}}\calH_{\rmf,00}(K_1,K_2)\int\limits_{K'\in\mathbb{R}^d}\mathrm{d}K'\int_{\eta=0}^{Z/2} \exp\Big(-\frac{i}{k}K'\cdot(K_1-K_2)\eta\Big)\mathrm{d}\eta\\ &\times\int_{\mu=-2\eta}^{2\eta}F(K',\mu)\hat{G}(K',\mu)\exp\Big(-\frac{i}{2k}K'\cdot(K_1+K_2)\mu\Big)\mathrm{d}\mu\\
&+\frac{k^2}{4(2\pi)^{d}}\calH_{\rmf,00}(K_1,K_2)\int\limits_{K'\in\mathbb{R}^d}\mathrm{d}K'\int_{\eta=Z/2}^{Z} \exp\Big(-\frac{i}{k}K'\cdot(K_1-K_2)\eta\Big)\mathrm{d}\eta\\ &\times\int_{\mu=-2(Z-\eta)}^{2(Z-\eta)}F(K',\mu)\hat{G}(K',\mu)\exp\Big(-\frac{i}{2k}K'\cdot(K_1+K_2)\mu\Big)\mathrm{d}\mu\\
\approx&\frac{k^2}{4(2\pi)^{d}}\calH_{\rmf,00}(K_1,K_2)\int\limits_{K'\in\mathbb{R}^d}\mathrm{d}K'\int_{\eta=0}^Z \exp\Big(-\frac{i}{k}K'\cdot(K_1-K_2)\eta\Big)\mathrm{d}\eta\\ &\times\int_{\mu\in\mathbb{R}}F(K',\mu)\hat{G}(K',\mu)\exp\Big(-\frac{i}{2k}K'\cdot(K_1+K_2)\mu\Big)\mathrm{d}\mu
\end{aligned}
\end{equation}
where the second estimate comes from~\eqref{eqn:assume_2}. The estimate includes the extra integration area $\int^{\delta_z/2}_{0}\int_{2\eta}^{\delta_z}\rd\mu\rd\eta\sim\mathcal{O}(\delta_z^2)\to0$ for small $\delta_z$.
% This is an integration of~\ql{how many dimension...? I actually don't see why it helps...}

% \sam{Sam is wondering: Why does it become $\int_{\mu\in\mathbb{R}}$? I was expecting it to become $\int_{\mu=-\delta_z}^{\delta_z}$, since those are the only $\mu$ values for which $F\neq 0$, right? Should we clarify why we want to write it as $\int_{\mu\in\mathbb{R}}$?  It it because you would like to carry out the $\mu$ integration in the next subsection?} \an{Writing the limits of integration that way helps to use the integral as a Fourier transform}

\subsection{Small-length-scale cutoff assumption}\label{sec:small_scale_cutoff_assumption}

In addition to the assumption from~\eqref{eqn:assume_2}, we now suppose the turbulence is characterized by the Kolmogorov model and with $l_0$ and $L_0$ being the inner and outer scales of turbulence, respectively. This means, if we write the power spectral density, the complete Fourier transform of the covariance function, given by:
\begin{equation}
\begin{aligned}
    \Phi(K,K_z)&=\epsilon^2 \int\limits_{z\in\mathbb{R}}\int\limits_{X\in\mathbb{R}^d}f(X,z)e^{-i(K\cdot X+K_zz)}\mathrm{d}X\mathrm{d}z\\
    &=\epsilon^2 \int\limits_{z\in\mathbb{R}}F(K,z)e^{-K_zz}\mathrm{d}z
\end{aligned}
\end{equation}
There are a few assumptions in place:

Firstly, we assume $\Phi$ is a power law in the inertial subrange $1/L_0\ll |(K,K_z)|=\sqrt{\|K\|^2+{K_z}^2}\ll 1/l_0$. Letting $L_0=\infty$, we run the von Karman spectrum multiplied by a Gaussian factor approximation outside the inertial subrange to ensure that the power spectrum decays rapidly for wavenumbers larger than $1/l_0$~\cite{andrews2005laser}. These assumptions makes $F(K,z)$ negligible for $\|K\|\gg 1/l_0$. 

Secondly, as noted in \cite{clifford1978classical}, we also assume $F(K, z)$ is negligible also for $\|K\|z>1$. These assumptions together suggest that $F(K,z)$ is not negligible only when:
\[
\frac{\|K\|^2z}{2}<\frac{\|K\|}{2}<\frac{1}{2l_0}\,.
\]

However, in this region, we recall the definition of $\hat{G}$ in~\eqref{A0:sol_Fourier}, we will see that $\hat{G}\sim 1$. This can be seen evaluating
\[
\hat{G} =\exp\{-\frac{i z}{2k}\|K\|^2\}\sim \exp(-\frac{i}{2kl_0})\sim\exp\{0\}=1\,,
\]
where we used the assumption that \revision{$1/(2kl_0)\ll 1$}. This assumption is realistic in the atmosphere for waves in the optical/IR regime according to ~\cite{andrews2005laser}. Plug this approximation back into equation~\eqref{eqn:H_f_11_dec_z}, we have:
% So in the region of interest, we can use the approximation that  $\|K'\|^2\mu/2k<\|K'\|z /2kl_0 \ll 1/2kl_0$. If $k>1/2l_0$, we have $\hat{G}(K',z)\approx 1$. So we have \ql{why $\mu<1$ and $\|K'\|l_0<1$? Do we need some justifications?}\an{some justification is given in Strohbehn chapter 2.}\an{For the following formulae to be true, we should be able to approximate $\hat{G}(K,z)=exp(-iK^2z/2k)$ by 1 in the region where $F(K,\mu)$ is non negligible. If $F$ has a correlation length of $l_0$ in $z$ direction and $2/l_0$ in $K$ direction, $F$ is negigible for $K>6/l_0$ or $\mu>3l_0$. So the region of interest is $K^2z/2k<36*3/2kl_0=54/kl_0$. So as long as $k\gg 54/l_0$, the approximation $\hat{G}\approx 1$ should be fine.}
\begin{equation}\label{eq:H_11_Fourier_new}
    \begin{aligned}
         &\calH_{\rmf,11}(K_1,K_2)\\
         \approx&\frac{Zk^2}{4(2\pi)^{d}}\calH_{\rmf,00}(K_1,K_2)\int\limits_{K'\in\mathbb{R}^d}\exp\Big(-\frac{i}{2k}K'\cdot(K_1-K_2)Z\Big)sinc\Big(\frac{K'}{2k}\cdot(K_1-K_2)Z\Big)\mathrm{d}K'\\
         &\times\int_{\mu\in\mathbb{R}}F(K',\mu)\exp\Big(-\frac{i}{2k}K'\cdot(K_1+K_2)\mu\Big)\mathrm{d}\mu\\
         =&\frac{Zk^2}{4(2\pi)^{d}\epsilon^2}\calH_{\rmf,00}(K_1,K_2)\int\limits_{K'\in\mathbb{R}^d}\exp\Big(-\frac{i}{2k}K'\cdot(K_1-K_2)Z\Big)sinc\Big(\frac{K'}{2k}\cdot(K_1-K_2)Z\Big)\\
         &\times\Phi(K',K'\cdot(K_1+K_2)/2k)\mathrm{d}K'\,.
    \end{aligned}
\end{equation}
Similarly, we have
\begin{equation}\label{eq:H_02_Fourier_new}
    \begin{aligned}
         \calH_{\rmf,02}(K_1,K_2)&\approx-\frac{Zk^2}{8(2\pi)^{d}\epsilon^2}\calH_{\rmf,00}(K_1,K_2)\int\limits_{K'\in\mathbb{R}^d}\Phi(K',-K'\cdot K_2/k)\mathrm{d}K'
    \end{aligned}
\end{equation}
and $\calH_{\rmf,20}=\calH^\ast_{\rmf,02}$.
\subsection{Markov approximation}\label{sec:Markov approximation}
One further simplifies the calculation when Markov approximation is imposed.

Let the covariance $f$ take the form of
\begin{equation}\label{eqn:cov_separable}
    f(x,z)=f_1(x)f_2(z)\,,
\end{equation}
then equations~\eqref{eqn:H11_Fourier_new} and~\eqref{eqn:H02_Fourier_new} get re-written as
\begin{equation}\label{eqn:Hf11_separable}
    \calH_{\rmf,11}=\frac{k^2}{4(2\pi)^{d}}\calH_{\rmf,00}(K_1,K_2)\int_{z_1=0}^Z\int_{z_2=0}^Zf_2(z_1-z_2)\widehat{F_1\hat{G}(\cdot}, z_1-z_2)\big((K_1z_1-K_2z_2)/k\big)\mathrm{d}z_2\mathrm{d}z_1\,,
\end{equation}
and
\begin{equation}\label{eqn:Hf02_separable}
   \calH_{\rmf,02}(K_1,K_2)=-\frac{k^2}{8(2\pi)^{d}}\calH_{\rmf,00}(K_1,K_2)\int_{z_1=0}^Z\int_{z_2=0}^{Z}f_2(z_1-z_2)\widehat{F_1\hat{G}(\cdot}, z_1-z_2)\big(K_2(z_2-z_1)/k\big)\mathrm{d}z_2\mathrm{d}z_1\,,
\end{equation}
where $F_1=\hat{f}_1$. Markov approximation means that $f_2=\delta(z)$, meaning the turbulence for every $z$ point is completely independent, then:
\begin{equation}\label{eqn:cov_Markov}
    f(x,z)=f_1(x)\delta(z)\,,
\end{equation}
which further reduces the computation of $\calH_{\rmf,11}$ and $\calH_{\rmf,02}$ to
\begin{equation}\label{eqn:Hf_11_Markov}
    \calH_{\rmf,11}=\frac{Zk^2}{4}\calH_{\rmf,00}(K_1,K_2)\int_{y=0}^1f_1\big((K_2-K_1)yZ/k\big)\mathrm{d}y\,,
\end{equation}
and
\begin{equation}\label{eqn:Hf_02_Markov}
    \calH_{\rmf,02}(K_1,K_2)=-\frac{Zk^2}{8}f_1(0)\calH_{\rmf,00}(K_1,K_2)\,,
\end{equation}
with $\calH_{\rmf,20}(K_1,K_2)=\calH_{\rmf,02}^\ast(K_1,K_2)$. Note that~\eqref{eqn:Hf_11_Markov} is a single integral and is computationally cheap.

We should note, however, when the assumption is this strong, the computation on the physical domain is similarly simple. Indeed, define $\eta=\frac{X_1'+X_2'}{2}$ and $\mu=X_1'-X_2'$, ~\eqref{eqn:H11} becomes: 
\begin{equation}\label{eqn:H11_separable}
\begin{aligned}
    \calH_{11}&=\frac{k^2}{4}\int\limits_{z_1=0}^Z\Big(\frac{k}{2\pi(Z-z_1)}\Big)^dG(X_1,z_1)G^\ast(X_2,z_1)\mathrm{d}z_1\int\limits_{\mu\in\mathbb{R}^d}f(\mu)\exp\big(-\frac{ik\mu(X_1+X_2)}{2z_1}\big)\mathrm{d}\mu\\
    &\times\int\limits_{X'\in\mathcal{R}}\exp\big(-\frac{ikX'\cdot\mu}{Z-z_1}\big)\mathrm{d}X'\int\limits_{\eta\in\mathbb{R}^d}\exp\Big(ik\eta\big(\frac{Z\mu-(X_1-X_2)(Z-z_1)}{z_1(Z-z_1)}\big)\Big)\mathrm{d}\eta\\
    &=\frac{Zk^2}{4}\calH_{00}(X_1,X_2)\int\limits_{y=0}^1f_1\big((X_1-X_2)y\big)\mathrm{d}y\,.
\end{aligned}
\end{equation}
Similarly,~\eqref{eqn:H02} becomes:
\begin{equation}\label{eqn:H02_separable}
    \calH_{02}=-\frac{Zk^2}{8}f_1(0)\calH_{00}(X_1,X_2)\,.
\end{equation}
This gives a one-fold integral and is numerically easy as well. This means when Markov approximation holds true, the computation on the physical domain is similarly easy with its counterpart from Fourier domain.

\section{Numerical examples}\label{sec:numerics}

% \sam{Anjali, do you need to add \emph{units} of m$^{-1}$ to the statement that $k=2\pi\times 10^6$? Is that needed in several places?}
% \an{I wonder if the unit is rad/m}
% \sam{Should we use a more precise term for $J$ instead of \emph{shape}? (And I understand that I probably used that term myself.) What is the most appropriate term? Mutual coherence function? Should search for the word ``shape'' and replace it.}

In this section, we present numerical examples to demonstrate the proposed methods for computing the optimal beam and its associated mutual intensity function.

To set up the computation, we assume a 2D domain, with the transmitter region being $X\in[-r,r]$ with $r=0.05$ m, and the receiver being located at $Z=3000$ m. The wave frequency is set to be $k=2\pi\times 10^6$ rad/m. For discretization we use a mesh size of $\Delta x=0.001$m. This gives 100 grid points in the $x$ direction within the transmitter region. The computational domain is $(x,z)\in[-L,L]\times[0,Z]$, with $L=1$ m. The domain half-width $L$ is set to be much larger than $r$ to ensure that the waves at the computational boundary are negligible. We use periodic boundary conditions, \revision{meaning $A(-L,z)=A(L,z)$, and $\partial_xA(-L,z)=\partial_xA(L,z)$} in the $x$ direction. From the perspective of Fourier space, the wave numbers range from $[-\pi/\Delta x,\pi/\Delta x)$ with $\Delta K=\pi/L$. This leads to 2000 grid points in Fourier space. 

In what follows, in subsection~\ref{subsec:physical-vs-fourier}, we compare $\calH$ computed in physical space, and $\calH_{\rmf}$ computed in Fourier space. It will be seen that the numerical results of $\calH$ and $\calH_{\rmf}$ are on top of each other, which suggests a fine enough resolution is being used. In subsection~\ref{subsec:numerics_optimal}, we present the shapes of optimal beams when different assumptions are incorporated. The results suggest that the small-length-scale cutoff assumption, which significantly reduces the numerical cost, provides accurate approximations in the calculation of $\calH$. In subsection~\ref{subsec:numerics_sensitive}, we show the behavior of the optimal beams under different levels of the strength of the turbulence.

\subsection{Validation of numerical resolution}\label{subsec:physical-vs-fourier}
\begin{figure}[htbp]
    \centering
    \includegraphics[width=10cm]{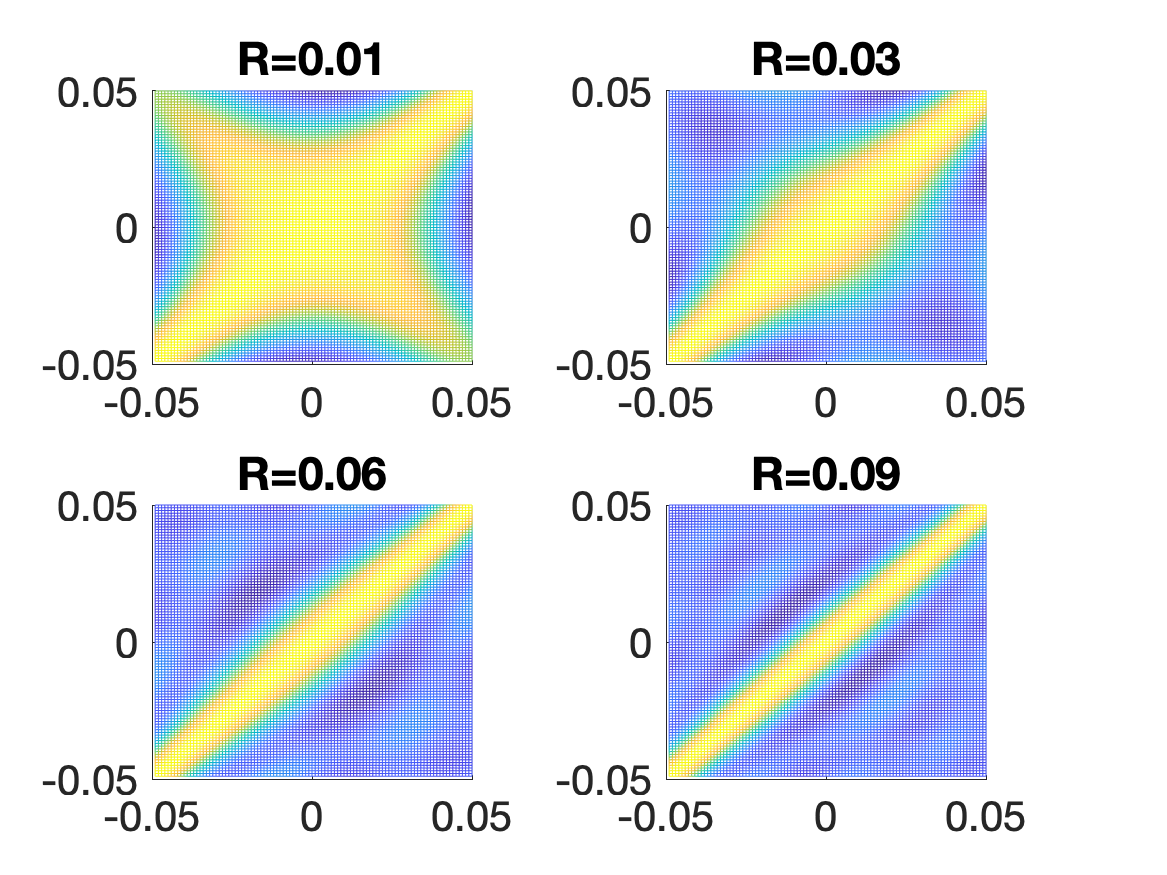}
    \caption{$\calH$ computed using~\eqref{eqn:H00_phys_space_simplification} when the receiver size $R$ is $0.01$, $0.03$, $0.06$, and $0.09$ m. The transmitter size is fixed at $r=0.05$ m.}
    \label{kernel_free_space}
\end{figure}

To get started, we first numerically verify the equivalence between the computation provided from the physical space~\eqref{eqn:H00} and that from the Fourier space~\eqref{eqn:H00_Fourier}. We assume $\epsilon=0$ so there is no turbulence in the medium. For a receiver region of $X\in\mathcal{R}=[-R,R]$ \revision{with $R$ being the radius of the receiver region}, we rewrite~\eqref{eqn:H00} to be:
% In the first examples here, a comparison is shown between
% the physical-space methods and Fourier-space methods as a
% validation test. For these validation tests,
% we assume that the medium has no turbulent fluctuations. 
% In this case, the medium is deterministic and $\epsilon=0$, and we calculate $\calH=\calH_{00}$ using~\eqref{eqn:H00}. Under the given setup, if the receiver region is given by $X\in[-R,R]$, we have
\begin{equation}\label{eqn:H00_phys_space_simplification}
    \calH_{00}(X_1,X_2)=\Big(\frac{Rk}{\pi Z}\Big)\exp\Big(\frac{ik}{2Z}(X_1^2-X_2^2)\Big)sinc\Big(\frac{Rk}{Z}(X_1-X_2)\Big),\quad (X_1,X_2)\in[-r,r]\,.
\end{equation}
Similarly, for this particular setup, we rewrite~\eqref{eqn:H00_Fourier} to be:
\begin{equation}\label{eqn:H00_Fourier_space_simplification}
    \calH_{\rmf,00}(K_1,K_2)=\frac{R}{\pi}\exp\Big(-\frac{iZ}{2k}(K_1^2-K_2^2)\Big)sinc\big(R(K_1-K_2)\big)\,.
\end{equation}
In Figure~\ref{kernel_free_space}, we plot $\calH$ with different $\mathcal{R}$, the size of the receiver. As shown in the plots, as $\mathcal{R}$ increases, $\calH$ becomes closer and closer to the identity matrix, with more and more eigenvalues closer to $1$. Physically this means that all modes from the transmitter arrive at the receiver with the intensity preserved.

% \sam{THIS IS WHERE SAM LEFT OFF ON JANUARY 9, 2022 !!!!! }

% \sam{Question from Sam: Why do we mention ``numerical integral'' below? For these cases, we have analytical formulas for $\calH$, right? So there are no numerical integrals needed, right?  Or do you consider the FFT to be the numerical integral?}

\begin{figure}[htbp]
    \centering
   \includegraphics[width=0.45\textwidth]{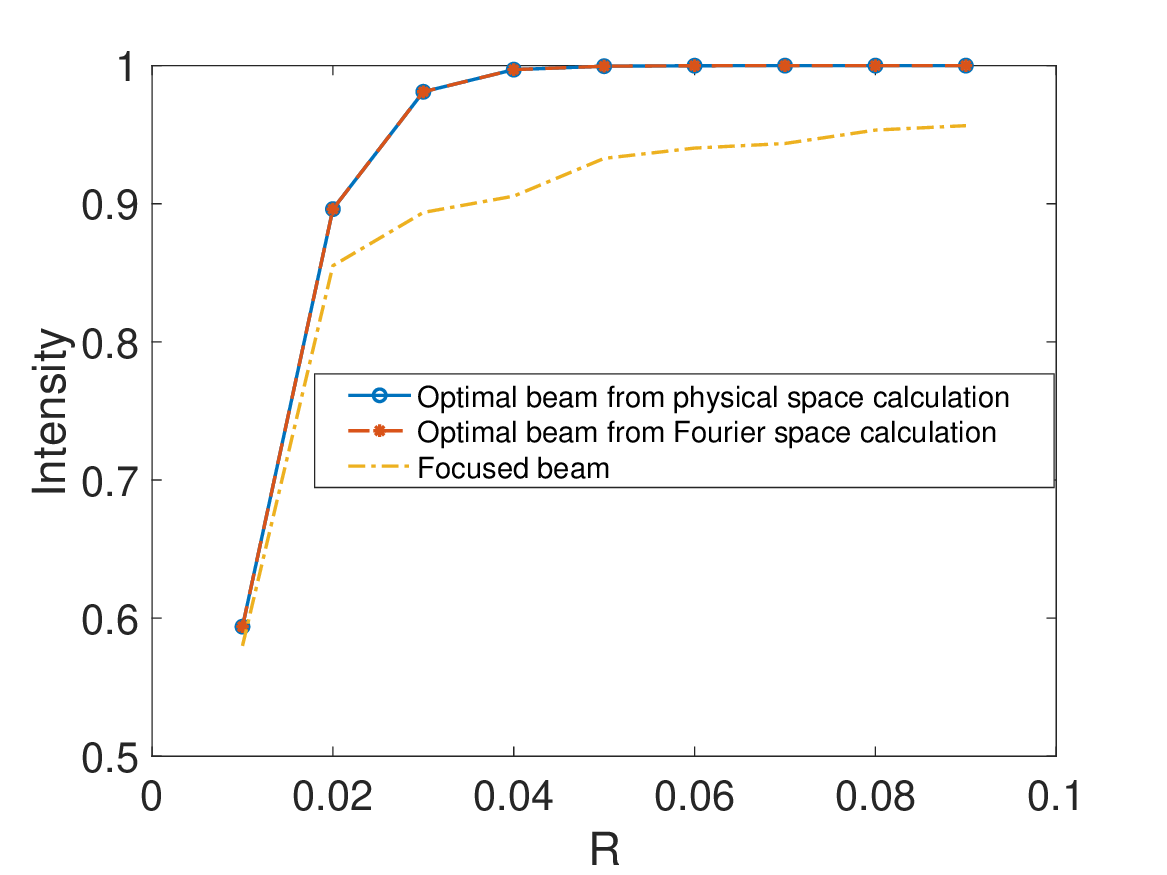}
\includegraphics[width=0.45\textwidth]{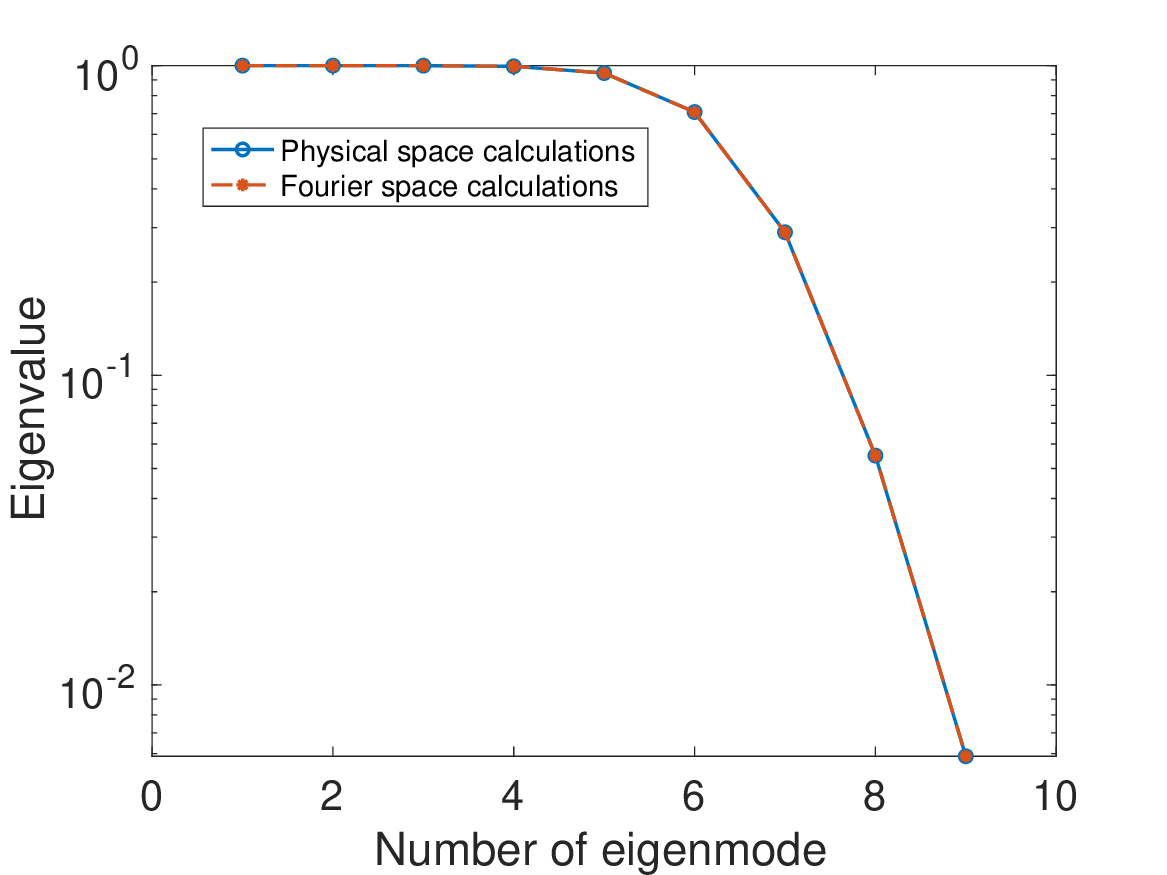}
    \caption{In the panel on the left, we show the comparison of total intensity over various receiver sizes, and the focused beam solution. Receiver size is in meters. In the panel on the right, we show the agreement of the first nine eigenvalues given by $\calH$ and $\calH_\rmf$, taking $R=0.09$m.}
    \label{Intensity_compare_free_space}  
\end{figure}

\begin{figure}[htbp]
\begin{subfigure}{0.32\textwidth}   
    \centering
    \includegraphics[width=4.5cm]{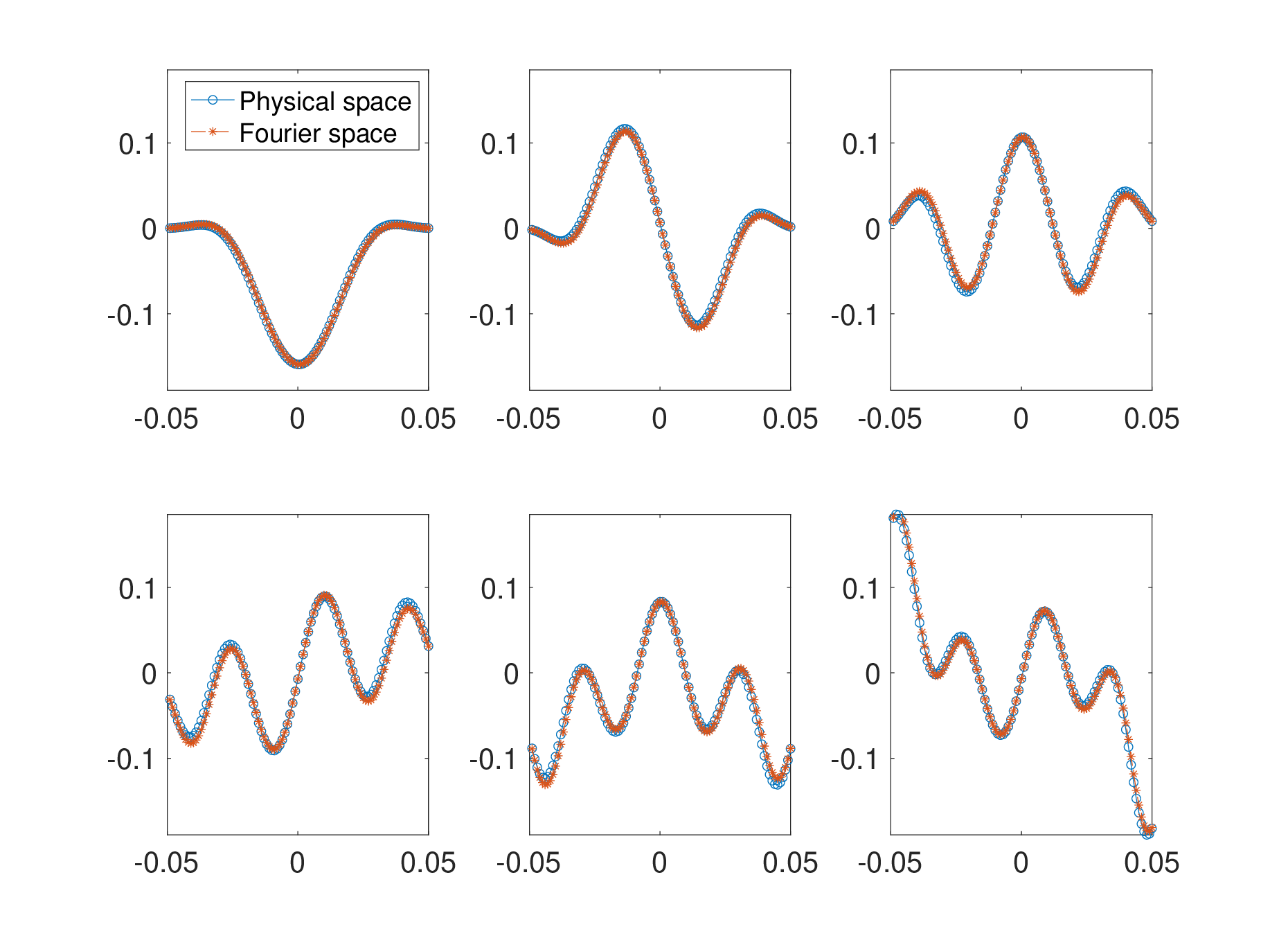}
    \caption{Real part}\label{fig:vector_free_real}
\end{subfigure}
 \begin{subfigure}{0.32\textwidth}    
    \centering
    \includegraphics[width=4.5cm]{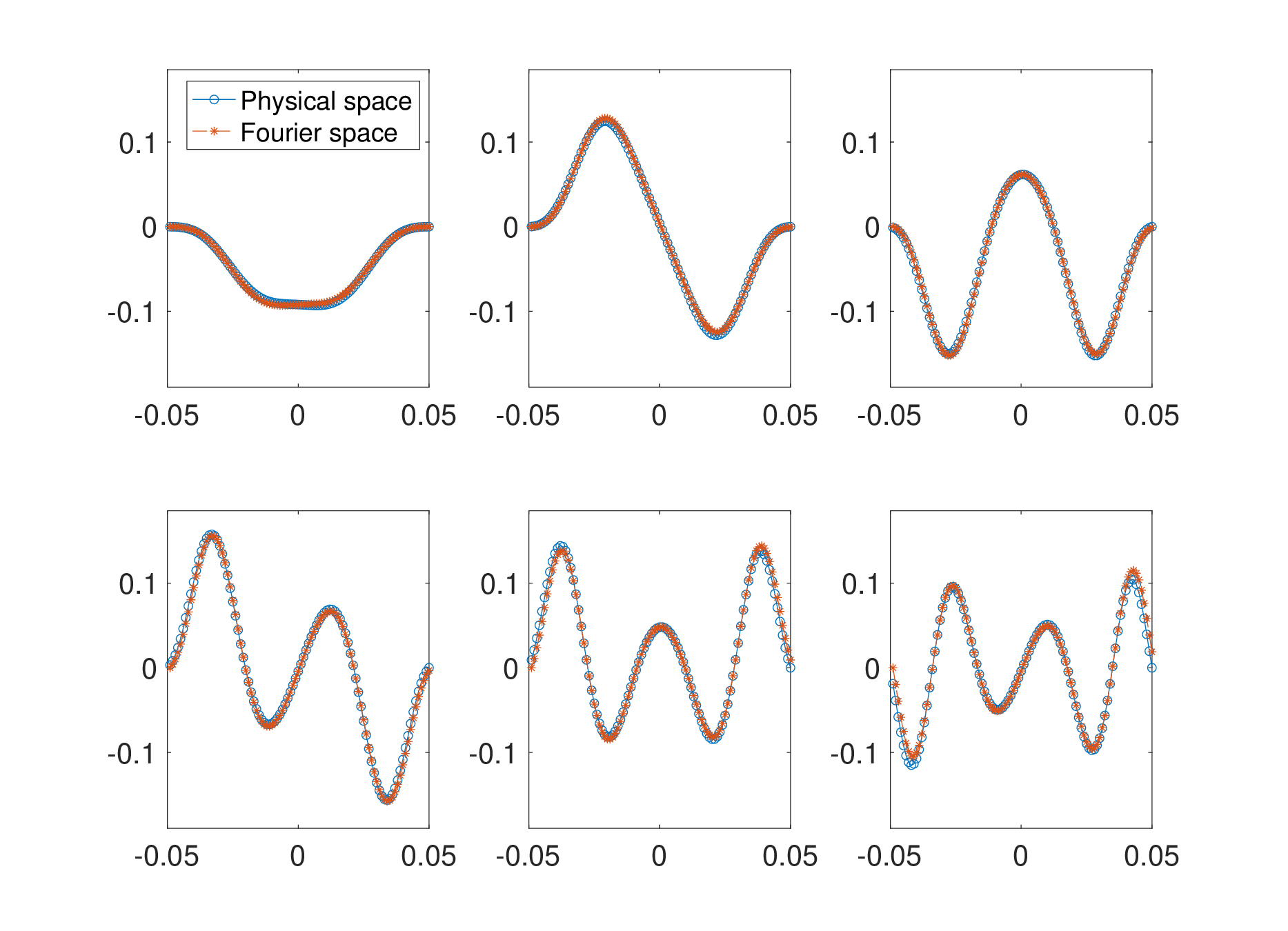}
    \caption{Imaginary part}\label{fig:vector_free_imag}
    \end{subfigure}
    \begin{subfigure}{0.32\textwidth}    
    \centering
    \includegraphics[width=4.5cm]{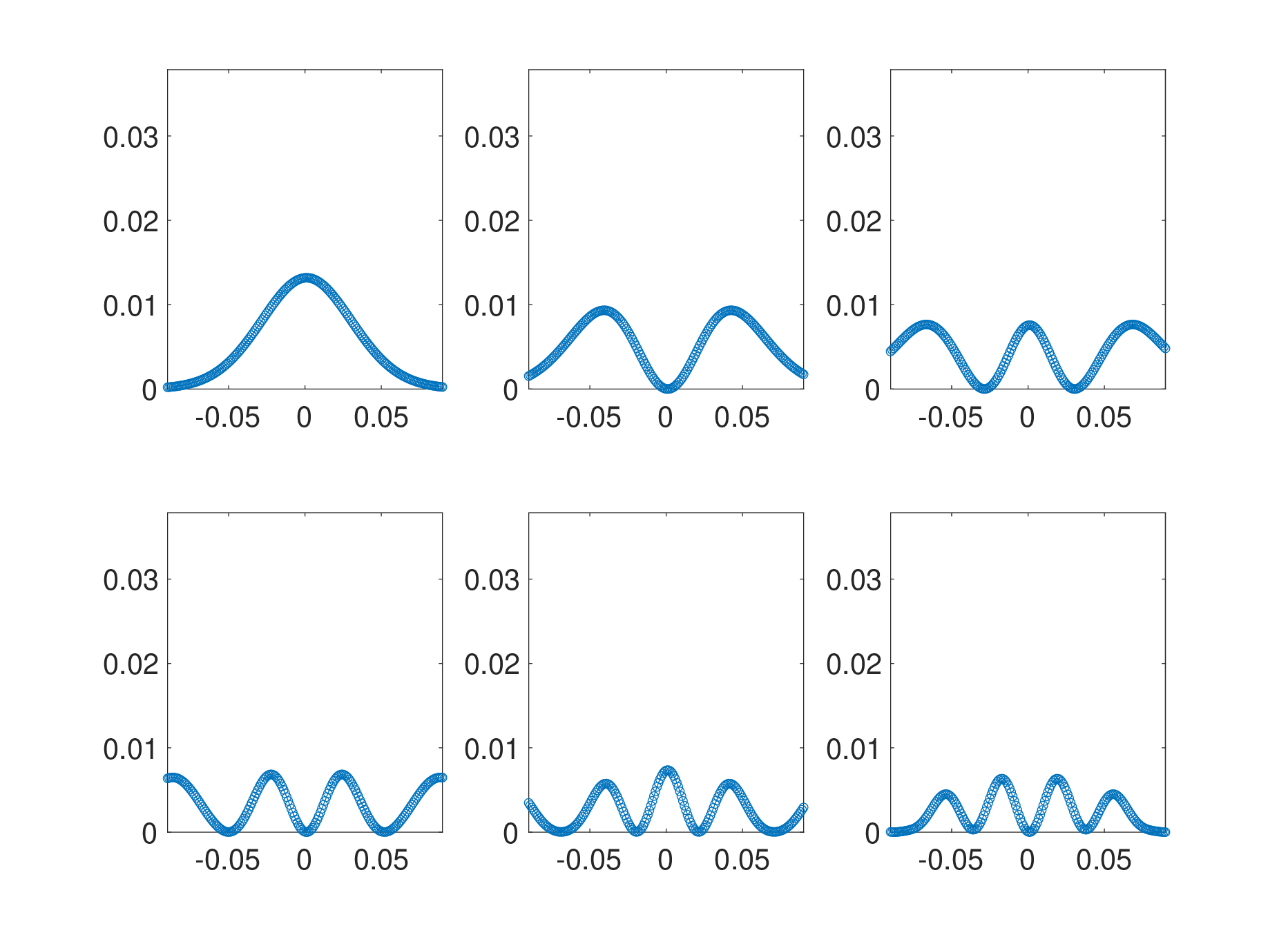}
    \caption{Intensity distribution at the receiver}
    \end{subfigure}
    \caption{Eigenfunctions provided by computing $\calH$ (blue line) and $\calH_{\rmf}$ (red line) are on top of each other. The left and the middle panel show the real and imaginary parts of the first six engenvectors, and the panel on the right shows the light intensity received at the receiver with the $R = 0.09 $m for each of the six eigenfunctions.}
    \label{vector_free_space}
\end{figure}

Computing the optimal beams using~\eqref{eqn:H00_phys_space_simplification} and~\eqref{eqn:H00_Fourier_space_simplification} should agree. This is shown in Figure~\ref{Intensity_compare_free_space} and Figure~\ref{vector_free_space}. In particular, in Figure~\ref{Intensity_compare_free_space} we demonstrate the intensity of the optimal beam for different $R$, and the agreement of the first nine eigenvalues. We should note that the optimal beams give higher intensity than the focused beam (using initial data $A(X,z=0)=\phi(X)=e^{-\frac{ik}{2Z}X^2}$ as a complex Gaussian) for all $R$, suggesting the focused beams are not optimal. \revision{One interesting phenomenon to be observed here is that the optimal beam achieves the full intensity when the receiver size is only about $0.05$m. This is the same size as the transmitter, indicating that the beam divergence is small. This observation resonates with the calculation shown in Figure 4 of~\cite{schulz2004iterative} where it suggests the full intensity can be captured when $R\sim \frac{2Z}{kr}$, which agrees with our computation.} In Figure~\ref{vector_free_space}, we plot the profile of the beams. Once again, the calculation given on the physical space and that given on the Fourier space agree with each other. These agreements suggest the numerical resolution is fine enough for the numerical experiments to be trusted.

\begin{figure}[htbp]
    \centering
   \includegraphics[width=0.5\textwidth]{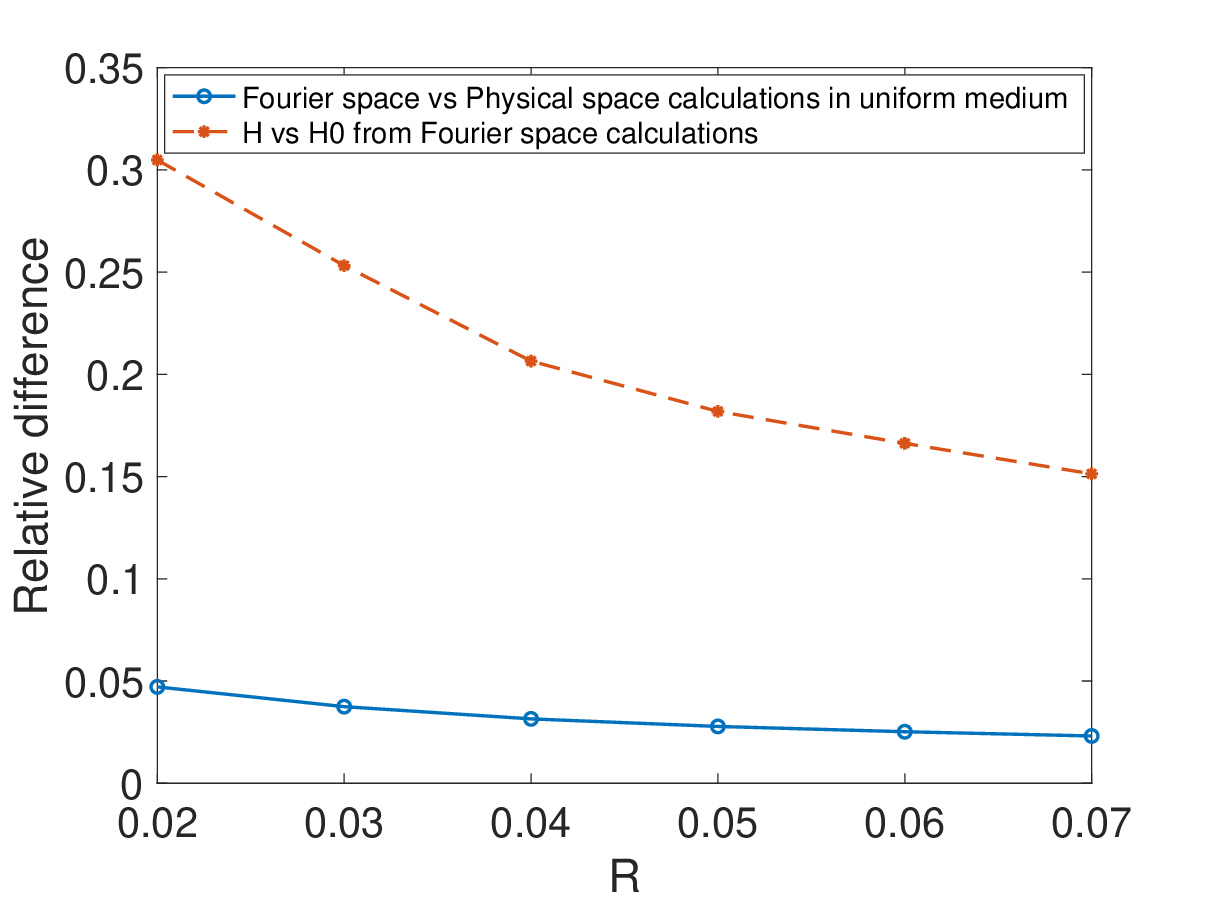}
    \caption{The plot in blue is the relative error in Fourier space vs physical space calculations for $\calH$ in a uniform medium (with respect to the Frobenious norm). The red plot shows the relative difference in the presence and absence of turbulence from Fourier space calculations. $\epsilon$ is fixed at $5\times 10^{-8}$ ($R$ is in meters)}
\label{fig:relative_error}
\end{figure}

In Figure~\ref{fig:relative_error}, we plot the difference of $\calH$ computed using the brute-force calculation~\eqref{eqn:H00_phys_space_simplification}, and that computed using the Fourier transform~\eqref{eqn:relation_H_f_p} from a simplified $\calH_\rmf$~\eqref{eqn:H00_Fourier_space_simplification}. For all $R$ this error is significantly smaller than the difference between $\calH$ computed for the different $\epsilon$ ($\epsilon=0$ vs. $\epsilon = 5\times 10^{-8}$, where further details of the $\epsilon>0$ case are described below). This is further evidence that the numerical errors are small compared to the changes in the solutions brought about by turbulent fluctuations.

%  \begin{figure}[htbp]
%     \centering
%     \includegraphics[width=10cm]{free_space_eigenvalue_comparison.png}
%     \caption{Comparison of eigenvalues from the two methods for aperture radius $0.05$ m and receiver radius $0.09 $ m}
%     \label{eigen_value_comparison_free_space}
% \end{figure}

%  \begin{figure}[htbp]
% \begin{subfigure}{0.5\textwidth}   
%     \centering
%     \includegraphics[width=7cm]{Intensity at transmitter_R=0.09.png}
%     \caption{Intensity distribution at the transmitter}
% \end{subfigure}
% \begin{subfigure}{0.5\textwidth}    
%     \centering
%     \includegraphics[width=7cm]{Intensity at receiver_R=0.09.png}
%     \caption{Intensity distribution at the receiver}
%     \end{subfigure}
%     \caption{Evolution of intensity profile of modes in free space designed for $r=0.05, R=0.09, Z=3000$ m.}
%     \label{intensity_compare_R=0.09}
% \end{figure}

\subsection{Shapes of optimal beams, and cost comparisons}\label{subsec:numerics_optimal}
\begin{figure}[htbp]
    \centering
    \includegraphics[width=0.45\textwidth]{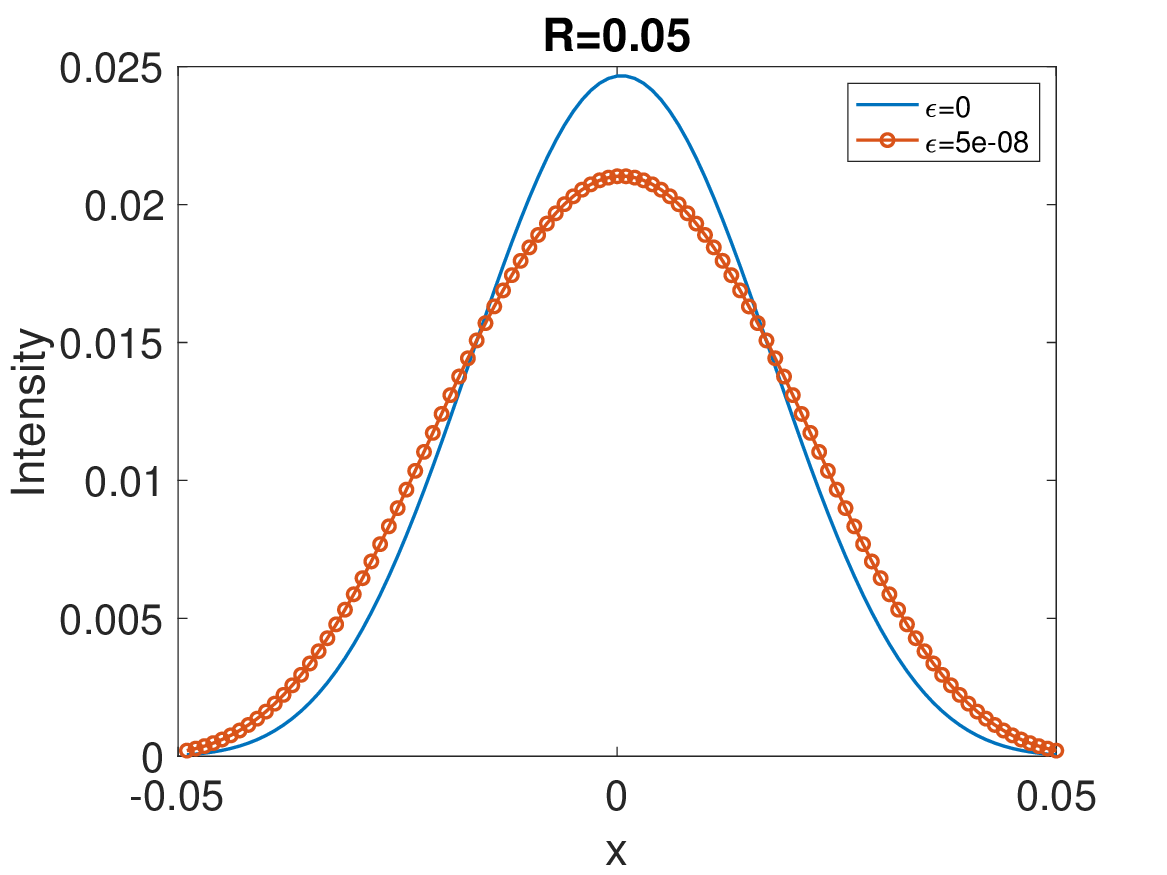}
   \includegraphics[width=0.45\textwidth]{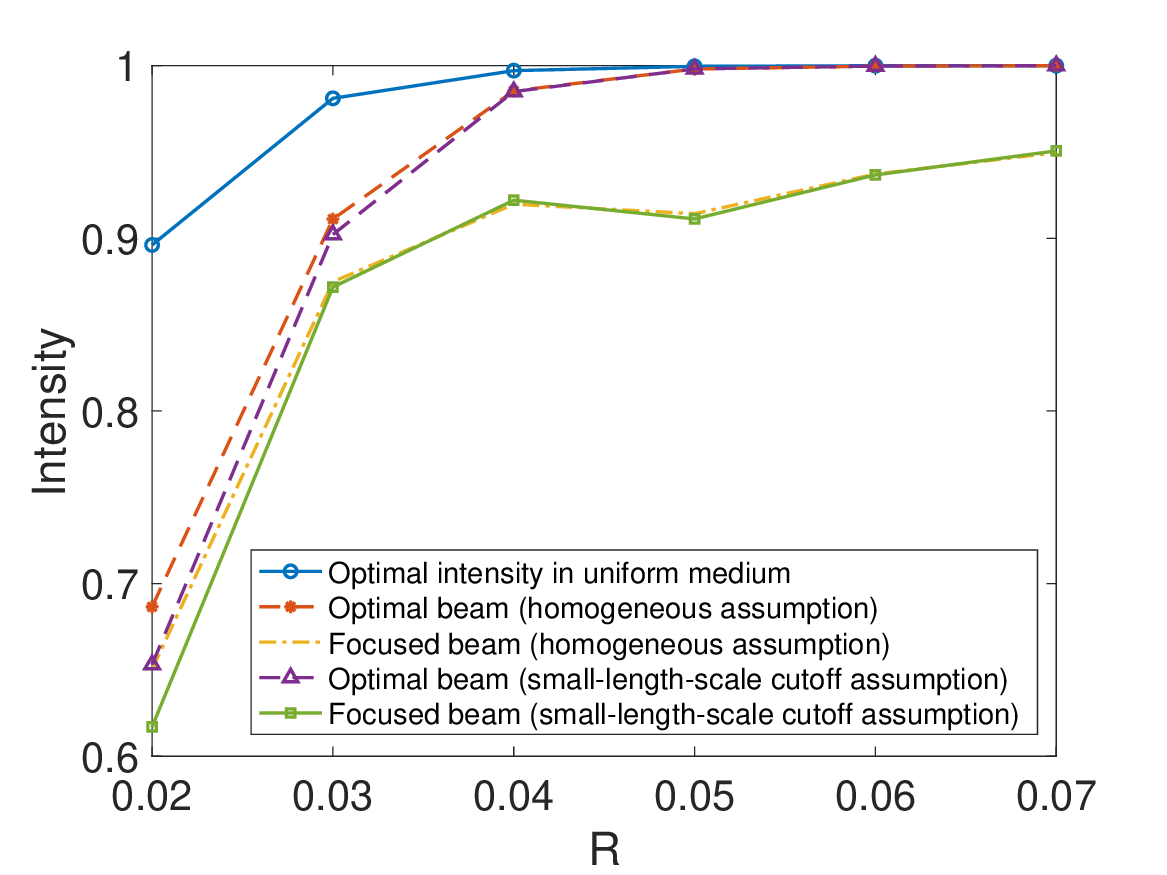}
    \caption{In the panel on the left, we show the {initial intensity profile} of optimal beam using $R=0.05$m. In the panel on the right, we show the comparison of total intensity at the receiver using optimal beams and focused beams under the homogeneous random medium assumption and small-length-scale cutoff assumption ($R$ is in meters).}
\label{fig:focused_comparison_intensity}
%     \caption{Absolute value of optimal beam for $R=0.05$ m}
% \label{fig:optimal_turb_R=0.05}
\end{figure}

In the second examples, we investigate cases with turbulent fluctuations in the medium, and we utilize the homogeneous assumption discussed in Section~\ref{sec:homog_assumption}. We use the same setup as above, and assume $\epsilon=5\times 10^{-8}$ (slightly larger than $0.3/k$). As a start, we assume $\Gamma$ to be a Gaussian function:
\begin{equation}
    \Gamma(X_1,z_1,X_2,z_2)=\exp\Big(-\frac{(X_1-X_2)^2+(z_1-z_2)^2}{l_0^2}\Big)
\end{equation}
with the correlation length $l_0=0.1$ m~\cite{andrews2005laser}. Under this assumption, we can compute $F$ to be:
\begin{equation}
    F(K,z)=\sqrt{\pi l_0^2}\exp\Big(-\frac{l_0^2K^2}{4}\Big)\exp\Big(-\frac{z^2}{l_0^2}\Big)\,.
\end{equation}
When the homogeneous assumption holds true, we take the formulas in~\eqref{eqn:H11_Fourier_new} and~\eqref{eqn:H02_Fourier_new} and insert
\begin{equation}
        \widehat{F\hat{G}}(K',z_1-z_2)=\sqrt{\pi l_0^2}\sqrt{\frac{\pi}{s(z_1-z_2)}} \exp\Big(-\frac{(z_1-z_2)^2}{l_0^2}\Big)\exp\Big(-\frac{K'^2}{4s(z_1-z_2)}\Big),
\end{equation}
where $s(z)=\frac{l_0^2}{4}+\frac{iz}{2k}$. Note that $\exp [-(z_1-z_2)^2/l_0^2]$ is negligible when $|z_1-z_2|>3l_0$, so one can perform integration in $z_1$ and $z_2$ in the domain of $(z_1,z_2)\in(0,Z]\times(z_1-3l_0,z_1)$ in the region of $z_2<z_1$. The same simplification can be used for $z_2>z_1$. With turbulence added in this way, we discover that the optimal beam is slightly wider than it is in the uniform-medium case, as illustrated in Figure~\ref{fig:focused_comparison_intensity} (left panel) computed using $R=0.05$m.

We furthermore test the accuracy of the small-length-scale cutoff assumption (see Section~\ref{sec:small_scale_cutoff_assumption}). For this purpose, we use~\eqref{eq:H_11_Fourier_new},~\eqref{eq:H_02_Fourier_new} with
\begin{equation}
    \Phi(K,K_z)=\epsilon^2\pi l_0^2\exp\Big(-\frac{l_0^2(K^2+K_z^2)}{4}\Big)\,.
\end{equation}
Noting that $\Phi(K,K_z)$ is negligible when $\|K\|>6/l_0$, we set $K\in[-K_{\max},K_{\max}]$ with $\Delta K=1/4l_0$ and $K_{\max}=6/l_0$. In Figure~\ref{fig:focused_comparison_intensity} (right panel), we test the light intensity given by the optimal beam using different assumptions in the calculation. The computation generated by using small-length-scale cutoff assumption agrees very well with that generated using the homogeneous assumption only, for all values of $R$. One should note, however, the computation using the homogeneous random medium assumption, as shown in equation~\eqref{eqn:H11_Fourier_new}, uses $2$-dimensional integral (for $d=1$ setup here), takes around $16$ minutes, while the same computation took around $3$ seconds using the small-length-scale cutoff assumption, where the computation is $1$-folded integral, suggested in~\eqref{eq:H_11_Fourier_new}. This holds true for every entry of $\mathcal{H}_\rmf$, and thus brings a significant savings in computation.

Also shown in Figure~\ref{fig:focused_comparison_intensity}
is a comparison of the optimal beam and the focused beam.
In all cases, the focused beam gives weaker light intensity at the receiver in comparison to the optimal beam. \revision{Also, while one might think that the optimization result is a trivial result because it has an approximately Gaussian profile of intensity, it is important to note that the optimal beam is complex-valued and the phase information is crucial. The complex-valued beam is not itself a Gaussian profile. See Figure~\ref{vector_free_space} for an example of a non-Gaussian, complex-valued beam profile which has a Gaussian profile of intensity.}

%the optimal beam profile does have a similar feature as a Gaussian function in the sense that it is a single localized bump. However, no assumptions on the beam profile were taken into consideration while performing the optimization. Furthermore, in the left panel of Figure~\ref{fig:focused_comparison_intensity} we compare the optimal beam with its corresponding Gaussian approximation. The profiles are different, suggesting that Gaussian profiles do not give the optimal intensity.

\revision{Figure~\ref{fig:focused_comparison_intensity} is evidence that the optimal beam has a beam divergence that is small. More specifically, note that the optimal beam has an intensity of approximately 1 after traveling a distance of $Z=3000$ m, from a transmitter of size $r=0.05$ m to a receiver of the same size ($R=0.05$ m). Hence, beam divergence must be small in order to allow the full intensity of the beam to reach such a small receiver. Such a phenomenon has been seen in previous calculations of optimal beams, in the case of a uniform non-turbulent medium and a phase screen model of turbulence \cite{schulz2004iterative}. In Figure~\ref{fig:focused_comparison_intensity}, this phenomenon is also seen in the case of the turbulence setup of the present paper.}

% \sam{Would it be good to remind the reader what calculations need to be done in order to get our results?  E.g., point to equations and say that an XX-dimensional integral needs to be computed?}

% we use a trapezoidal method with $z_1\in(0,Z]$ in steps of $\Delta z_1=20$ m and $z_2\in[z_1-3l_0,z_1)$ in steps of $\Delta z_2=l_0/8$ in the region where $z_1>z_2$.  The region $z_1<z_2$ can be re-written in a similar manner by a change of variables $(z_1,z_2)\to (z_2,z_1)$. We use $\epsilon=5\times 10^{-8}$, which is slightly larger than the magnitude of $0.3/k$. The profile of the absolute value of the optimal beam is plotted in Figure~\ref{fig:optimal_turb_R=0.05} for $R=0.05$ m. As we can see, the optimal beam is less concentrated when the media perturbation $\epsilon$ is nontrivial.

% The optimal intensity from both these assumptions are compared in Figure~\ref{fig:focused_comparison_intensity} along with the focused beam for different receiver sizes. The intensity of the optimal beam drops slightly for smaller receiver sizes and saturates to 1 as the receiver size increases. However, the intensity of the focused beam is smaller than that of the optimal beam even for large values of $R$.

% \begin{figure}[htbp]
%     \centering
%   \includegraphics[width=10cm]{Focused_beam_comparison.png}
%     \caption{Comparison of total intensity at the receiver using optimal beams and focused beams under the homogeneous random medium assumption and small-length-scale cutoff assumption ($R$ is in meters).}
% \label{fig:focused_comparison_intensity}
% \end{figure}

\subsection{Sensitivity studies}\label{subsec:numerics_sensitive}
\begin{figure}[htbp]
    \centering
   \includegraphics[width=10cm]{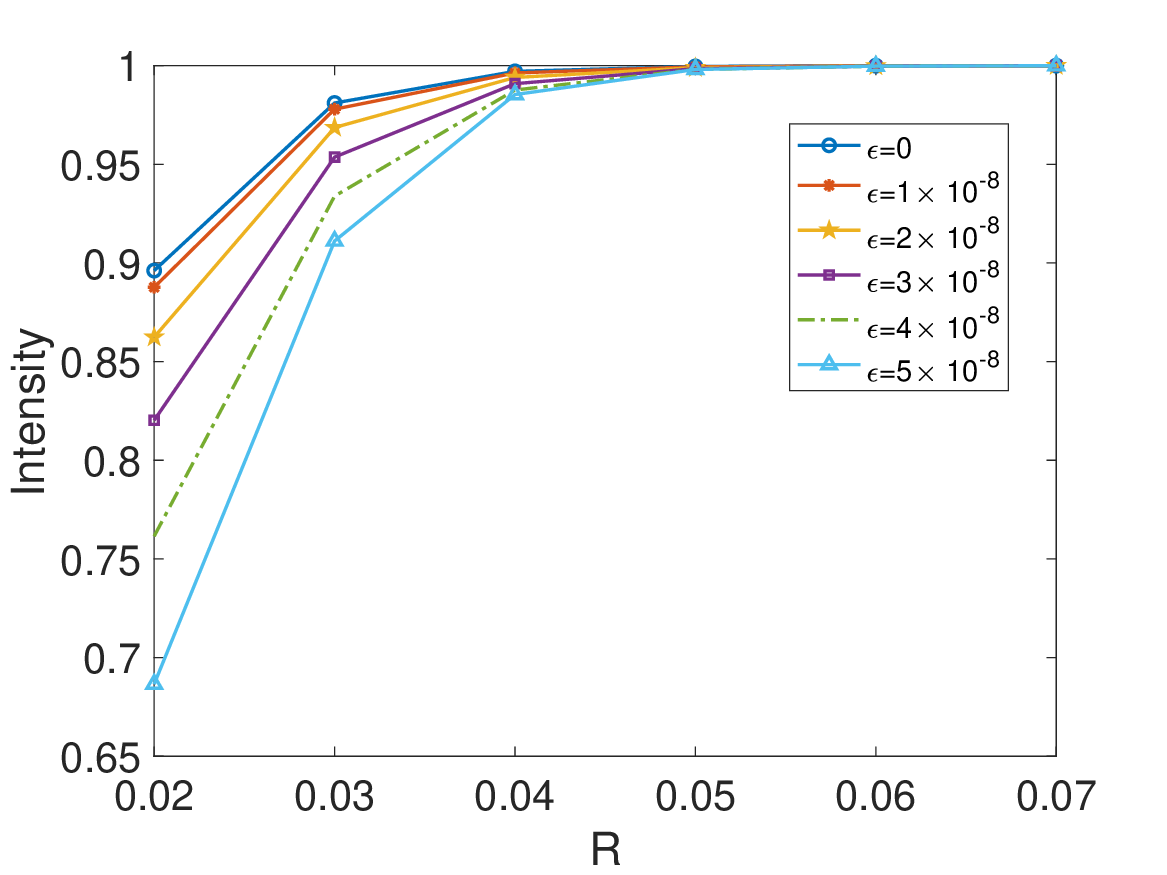}
    \caption{Optimal intensity at the receiver as $\epsilon$ increases ($R$ is in meters)}
\label{fig:epsilon_comparison_intensity}
\end{figure}

\begin{figure}[htbp]
    \centering
   \includegraphics[width=12cm]{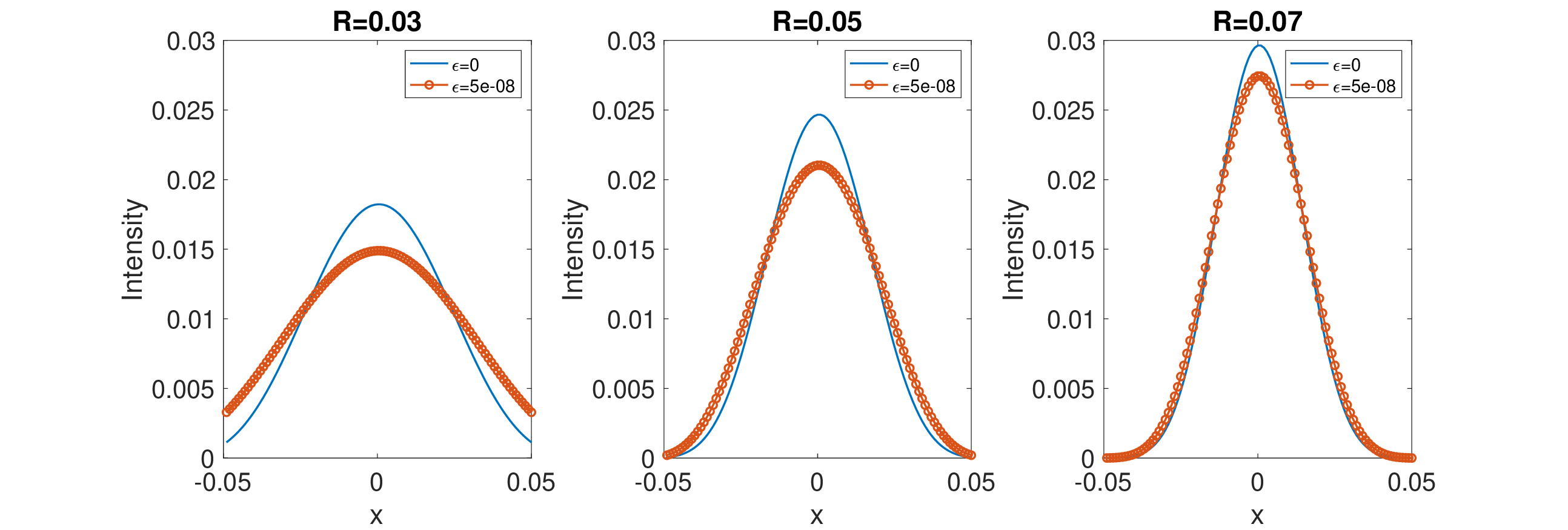}
    \caption{Initial intensity profile of optimal beams for different receiver sizes for $\epsilon=5\times 10^{-8}$ ($R$ is in meters)}
\label{fig:optimal_beams_turbulence}
\end{figure}

In this subsection, we study the relation between the optimal beam and the intensity of the turbulence. We would like to see the changes in the optimal beam that arise as greater turbulent fluctuations are introduced. As seen in Figure~\ref{fig:epsilon_comparison_intensity}, the total intensity of optimal beam drops as $\epsilon$ increases, for all choices of $R$. For small $R$, this drop is significant. The profile of the optimal beam also changes as $\epsilon$ changes, and the change is most prominent for small $R$, as shown in Figure~\ref{fig:optimal_beams_turbulence}.

% the performance of optimal beams for different turbulence strengths and different receiver sizes. We use the same setup as that of the homogeneous random medium assumption above. Figure~\ref{fig:epsilon_comparison_intensity} plots the total intensity of optimal beams as $\epsilon$ changes from $0$ to $5\times 10^{-8}$ for different receiver sizes. The optimal intensity drops as $\epsilon$ increases, but increases as the receiver size $R$ increases. The profile of the absolute value of the optimal beams is plotted in Figure~\ref{fig:optimal_beams_turbulence} for $\epsilon=5\times 10^{-8}$. In Figure~\ref{fig:relative_error}, we plot the relative error in Fourier space computations of $\calH$ compared to the physical space computations in free space (wrt Frobenious norm). We also plot the relative difference in the Frobenious norm of $\calH$ in turbulence compared to free space from Fourier space calculations. This relative difference is much larger than the relative error, which is very small. $\epsilon$ is fixed at $5\times 10^{-8}$.

\section{Concluding discussion}
In this paper, we computed profiles of optimal beams that achieve the highest intensity at the receiver. Mathematically this amounts to solving for eigenfunctions of $\calH$. In most realistic settings, $\calH$ is numerically infeasible to compute, with each entry calling for an $8$-fold integral. We proposed to convert the calculation to the Fourier domain where assumptions on the turbulent medium can be naturally incorporated. By introducing assumptions of spatially homogeneous statistics of the random medium, small-length-scale cutoff assumption, and Markov assumption, the $8$-fold integral is replaced by $6$-fold, $2$-fold and $1$-fold integrals respectively, with the numerical cost significantly reduced. This research generalizes the existing literature that concerns mostly the Markov approximation, or special classes of mutual intensity functions, and now allows for a general beam profile in a much more general turbulent medium structure. The methods proposed here point toward the possibility of computing general optimal beams.

\revision{The numerical examples here suggest that optimal beams can have nearly the full intensity transmitted, and small beam divergence. Similar results had also been seen in past studies under different numerical experimental setups \cite{schulz2004iterative}. These types of results show the influence of initial beam properties on the downstream characteristics of the beam.}

% \revision{LET'S CONSIDER ADDING SOME SENTENCES HERE ABOUT THE WAY THAT THE OPTIMAL BEAMS HAVE LESS BEAM DIVERGENCE THAN GAUSSIAN BEAMS. SCHULTZ HAD SHOWN THE SAME THING, ALTHOUGH WE NEED TO LOOK INTO THE DETAILS OF HIS SETUP TO SEE HOW GENERAL IT WAS. SAM ALREADY ADDED A SENTENCE ON THIS TOPIC TO THE ABSTRACT. MAYBE ALSO GOOD TO BRING UP THIS TOPIC IN THE RESULTS SECTION WHEN WE SHOW THE FIGURES?}

\section*{Acknowledgments}

The authors thank Svetlana Avramov-Zamurovic 
for helpful comments and discussion.
The research of Q.L. is partially supported by 
Office of Naval Research (ONR) grant N00014-21-1-2140,
and the research of A.N. and S.N.S. is partially supported by
ONR grant N00014-21-1-2119. 

\section*{Disclosures}
The authors declare no conflicts of interest.

\section*{Data availability}
Data and code underlying the results presented in this paper are available in~\cite{Nair2022}.

%%%%%%%%%% If using BibTeX:
\bibliography{Reference}

\end{document}